\documentclass[
 aip,
 reprint,
 jcp,
 nolongbibliography
]{revtex4-2}

\usepackage{graphicx}
\usepackage{bbm}
\usepackage{dcolumn}
\usepackage{amsmath,amssymb,bm}
\usepackage{mathtools}
\usepackage[bookmarks=false,breaklinks=true]{hyperref}
\usepackage{xcolor}
\usepackage[subrefformat=parens]{subcaption}
\usepackage{accents}
\usepackage{booktabs}
\newcommand{\vect}[1]{\boldsymbol{\mathbf{#1}}}

\newcommand{\angstrom}{\text{\normalfont\AA }}

\usepackage[normalem]{ulem}

\hypersetup{
  pdfstartview = FitH,
  pdftoolbar   = false,
  pdfmenubar   = true,
  colorlinks   = true,
  linkcolor    = blue!50!black,
  urlcolor     = blue!50!black,
  citecolor    = blue!50!black
}

\begin{document}

\preprint{AIP/123-QED}

\title{Diffusive dynamics of a model protein chain in solution}
\author{Margarita~Colberg}
\email{margarita.gladkikh@mail.utoronto.ca}
\author{Jeremy~Schofield}
\email{jeremy.schofield@utoronto.ca}
\affiliation{Chemical Physics Theory Group, Department of Chemistry,
University of Toronto, Toronto, Ontario M5S 3H6, Canada}

\date{\today}

\begin{abstract}
A Markov state model is a powerful tool that can be used to track the evolution
of populations of configurations in an atomistic representation of a protein.
For a coarse-grained linear chain model with discontinuous interactions, the
transition rates among states that appear in the Markov model when the monomer
dynamics is diffusive can be determined by computing the relative entropy of
states and their mean first passage times, quantities that are unchanged by the
specification of the energies of the relevant states. In this paper, we verify
the folding dynamics described by a diffusive linear chain model of the crambin
protein in three distinct solvent systems, each differing in complexity: a
hard-sphere solvent, a solvent undergoing multi-particle collision dynamics, and
an implicit solvent model. The predicted transition rates among configurations
agree quantitatively with those observed in explicit molecular dynamics
simulations for all three solvent models. These results suggest that the local
monomer-monomer interactions provide sufficient friction for the monomer
dynamics to be diffusive on timescales relevant to changes in conformation.
Factors such as structural ordering and dynamic hydrodynamic effects appear to
have minimal influence on transition rates within the studied solvent densities.
\end{abstract}

\maketitle

\section{INTRODUCTION} \label{sec:intro}

The study of the dynamics of protein folding {\it in silico} is challenging not
only due to the complexity and size of biomolecular systems but also because of
the separation of timescales between the dynamics of molecular motion and the
timescales of the significant conformational motions that are important to their
function. For atomistic models in which the constituents interact via molecular
mechanical force fields, the potential energy landscape is typically rugged,
characterized by many local minima corresponding to multiple metastable
molecular configurations in which the system remains for extended periods. In
brute force molecular dynamics (MD) simulations, the equations of motion must be
numerically integrated with a small time step to resolve motion on the
femtosecond timescale and conserve energy, while the conformational changes in a
protein folding process occur on the order of nanoseconds or
longer\cite{Lane:2013,Hollingsworth:2018}. Therefore, millions to billions of
time steps are required for an MD simulation to witness even a single major
structural change\cite{Lane:2013,Chodera:2014}. Such timescales are inaccessible
even with supercomputers such as Anton\cite{Shaw:2008,Chodera:2014}.

Markov state models (MSMs) have emerged as a popular approach to bridge this
timescale gap by predicting long timescale dynamics based on numerous short MD
trajectories\cite{Noe:2009,Pande:2010,Wang:2018/1,Husic:2018/2386,Konovalov:2021}.
In an MSM, the conformational space is partitioned into metastable states, such
that intrastate transitions are fast but interstate changes are slow. The
dynamics of populations in the targeted states over long times are predicted by a
Markovian master equation governed by a matrix of transition rates among them.
The discovery of the appropriate number and specification of the metastable
states, a critical component of the construction of an MSM, has attracted much
attention, and recent advances in machine learning methods have been used to
facilitate these tasks\cite{Konovalov:2021,Mardt:2022}. After identifying the
states, the transition rate matrix must be estimated using MD trajectories among
any dynamically connected states\cite{Bowman:2009,Noe:2009,Prinz:2011,Noe:2015}.
Once this transition matrix is available, the MSM can be analyzed to predict
several interesting properties. For example, the probability of particular
mechanistic paths starting from an initial distribution of states to the folded
state can be studied to find dominant folding pathways and potential bottlenecks
in the non-equilibrium folding process\cite{vonKleist:2018,Sharpe:2021}.
However, because the computed rate matrix is based on explicit simulations of
the system, it only applies to the conditions under which the trajectories were
performed, such as the temperature and the specific force field. As such, the
study of how temperature or mutations in sequence might change the folding
dynamics and its pathways requires reidentification of the target states and
reconstruction of the transition rate matrix.

Markov state models of simple coarse-grained protein systems can bypass these
restrictions to address how interaction energies should be chosen to optimize
specific dynamical properties and to elucidate how folding pathways for a given
model change with temperature\cite{Colberg:2022}. Additionally, detailed
questions concerning the nature of the free energy landscape can be
analyzed\cite{Bayat:2012/245103,Schofield:2014/095101}. If the bonds in the
secondary or tertiary structure of the protein are modeled using attractive step
potentials while all other interactions are taken to be hard-core repulsions,
the point at which bond formation or breaking occurs is defined geometrically.
The configurational space can be partitioned into non-overlapping regions, each
corresponding to a particular configuration or bonding pattern in the protein. A
dimensionless entropy can be determined for each configuration, dependent only
on the distances between the protein's monomers and independent of its energy.
Using this entropy, the canonical probability of a configuration can be easily
determined for any temperature and bonding energy. Consequently, the impact of
the choice of temperature and bonding energies in the model on measures of the
roughness of the free energy landscape can be examined\cite{Bayat:2012/245103}.

Previously\cite{Colberg:2022}, the dynamics of a protein-like chain were modeled
using an MSM in which the elements of the transition rate matrix were predicted
from a kinetic model that assumes diffusive monomer dynamics. In the simple
model, the rates of transitions between states, defined in terms of the
distances of a set of nonlocal contacts, can be computed primarily from
geometrical information, namely, the configurational entropy and the mean
squared diffusion distance between states, which are readily obtained from
importance sampling calculations that do not require simulating the real
dynamics. In Ref.~(\onlinecite{Colberg:2022}), importance sampling was performed
using hybrid Monte Carlo\cite{Duane:1987} based on artificial event-driven
dynamical trajectories. When bonding interactions are in the form of step
potentials, the diffusion coefficient along a bonding reaction coordinate,
needed to scale the mean squared diffusion distance to compute mean first
passage times, can also be computed from the sampling trajectories. Since the
kinetic model provides analytical expressions for the transition rate matrix in
terms of these quantities and the choice of state energies relative to the
temperature, dynamical characteristics of the protein-like chain, such as the
mean transition time between extended and native-like configurations, can be
predicted and optimized with respect to the choice of state energies.

The kinetic model, which underlies predictions, is based on a number of
assumptions that can be tested by comparing the predicted dynamics to those
observed in explicit simulations. This comparison can be accomplished by
estimating the transition rate matrix using any of the methods applied to
extract rates from MD trajectories of systems interacting with continuous force
fields\cite{Noe:2009,Prinz:2011,Noe:2015}. In this work, the predicted dynamics
of a diffusive MSM of a linear chain model of the crambin protein suspended in a
solvent bath are compared to those observed in molecular dynamics simulations.
The solvents considered here interact with the monomers of the chain via either
hard or effective interactions that do not alter the configurational entropy or
the probability of configurations. Since the simulation of the dynamics in the
presence of many solvent particles is computationally expensive, the dynamics of
selected transitions in the crambin model are analyzed. The transition rate
matrix elements for the selected states are determined for three solvent models
of differing complexity: one implicit and two explicit. In the implicit solvent
model, called the penetrating solvent, the presence of the solvent is mimicked
by a periodically applied stochastic random force to the monomers. This solvent
is the least computationally expensive, and the method is essentially equivalent
to Brownian dynamics simulations for event-driven dynamics. The penetrating
solvent provides a stochastic environment for the chain to allow thermal
equilibration but does not include hydrodynamic flow. The two explicit solvents
are modeled using multi-particle collision dynamics (MPCD) with hard
interactions and a hard-sphere fluid. Both solvent models allow for hydrodynamic
flow, and the latter takes into consideration the structure of the solvent
around the chain.

We find that the dynamics predicted by the diffusive MSM are in good agreement
with the dynamics observed in all three solvent models, even at low densities
with relatively large monomer self-diffusion coefficients. These results suggest
that the internal friction arising from the elastic collisions of monomers due
to excluded volume and local geometric constraints that maintain the linear
chain provides sufficient dissipation, when combined with weak solvent
collisions, to establish a separation of timescale between the timescale of bond
forming and breaking events and the decorrelation time of bead velocities.
Furthermore, for the solvent densities studied here, hydrodynamic flow and
solvent structure are not significant in determining the transition rates.

The outline of this paper is as follows: In Sec.~\ref{sec:cgm}, we introduce the
coarse-grained model, followed by a description of the specific interactions
chosen to represent crambin in Sec.~\ref{ssec:crambin}. The configurational
entropy and the canonical probability of macrostates are briefly discussed in
Sec.~\ref{ssec:thermo}, after which the transition rate matrix elements of the
diffusive MSM and the mean first passage times are presented in
Sec.~\ref{ssec:msm}. In Sec.~\ref{ssec:layer}, the layer method is briefly
reviewed, which is an approach developed previously\cite{Colberg:2022} to
compute the rate of transitions between intermediate states in a massively
parallel way. Sec.~\ref{ssec:diffcoeff} concludes this section by discussing the
diffusion coefficient needed to evaluate the mean first passage times. The three
solvent bath models are introduced in Sec.~\ref{sec:solventbath}, followed by a
discussion of the simulation parameters and their values in
Sec.~\ref{sec:simparam}. In Sec.~\ref{sec:simulations}, ten representative
transitions between crambin's intermediate states are selected, and the method
by which the diffusion coefficient for each transition was extracted from the
simulations is outlined in Sec~\ref{ssec:simdiffcoeff}. The equilibrium
probability and decay rate for crambin suspended in each of the three solvent
baths were obtained and compared to the canonical probability and transition
rate predicted by the diffusive MSM. Finally, results and concluding remarks are
provided in Secs.~\ref{sec:results} and~\ref{sec:concl}, respectively.
Additional background on the derivation of the diffusive Markov state model is
presented in the Appendix.

\section{THE COARSE-GRAINED MODEL OF PROTEINS} \label{sec:cgm}

The system of interest comprises a coarse-grained protein-like chain immersed in
a fluid in thermal equilibrium at temperature $T$. Each bead in the linear
chain in the model represents each amino acid in the primary structure, and
connective bonds between the beads are maintained using a square-well potential,
\begin{equation} \label{eq:squarewell}
    U_{ij} \left(r\right) =
    \begin{cases}
      0 & \text{if}\ \sigma_1 < r < \sigma_2 \text{,} \\
      \infty & \text{otherwise,}
    \end{cases}
\end{equation}
where $r$ is the distance between the nearest or next-nearest neighboring beads,
and $\sigma_1$ and $\sigma_2$ define the minimum and maximum bonding distances.
For the nearest neighbors in the chain, $\sigma_1 = 1$, which is defined to be
the unit of length $\ell$ in the model, and $\sigma_2 = 1.17$, whereas for the
next-nearest neighbors, $\sigma_1 = 1.4$ and $\sigma_2 = 1.67$ to confine the
bond angles to the range $75^{\circ}$--$112^{\circ}$. Nonlocal bonding
interactions, which can occur between monomers separated by more than two beads,
are modeled using a square-well potential,
\begin{equation} \label{eq:attractivestep}
    U_{ij} \left(r\right) =
    \begin{cases}
      \infty & \text{if}\ r < r_h \\
      \epsilon_{ij} & \text{if}\ r_h \leq r \leq r_c \\
      0 & \text{if}\ r > r_c.
    \end{cases}
\end{equation}
In Eq.~\eqref{eq:attractivestep}, when the beads $i$ and $j$ are separated by a
distance of $r < r_c = 1.5$, a nonlocal bond is formed between them, which
lowers the potential energy $U_{ij}$ by an amount $\epsilon_{ij} < 0$ relative
to the unbonded energy and stabilizes the overall system in the process.
Additionally, all nonlocal monomers interact via a hard-sphere repulsion with a
hard-sphere diameter of $r_h = 1.25$ to account for excluded volume interactions
at short distances.

\subsection{The crambin model} \label{ssec:crambin}

\begin{figure*}[htbp]
    \subfloat[
        \label{fig:crambinPDB}
        Crambin crystal structure from the Protein Data Bank
    ]{%
        \includegraphics[height=3.5cm]{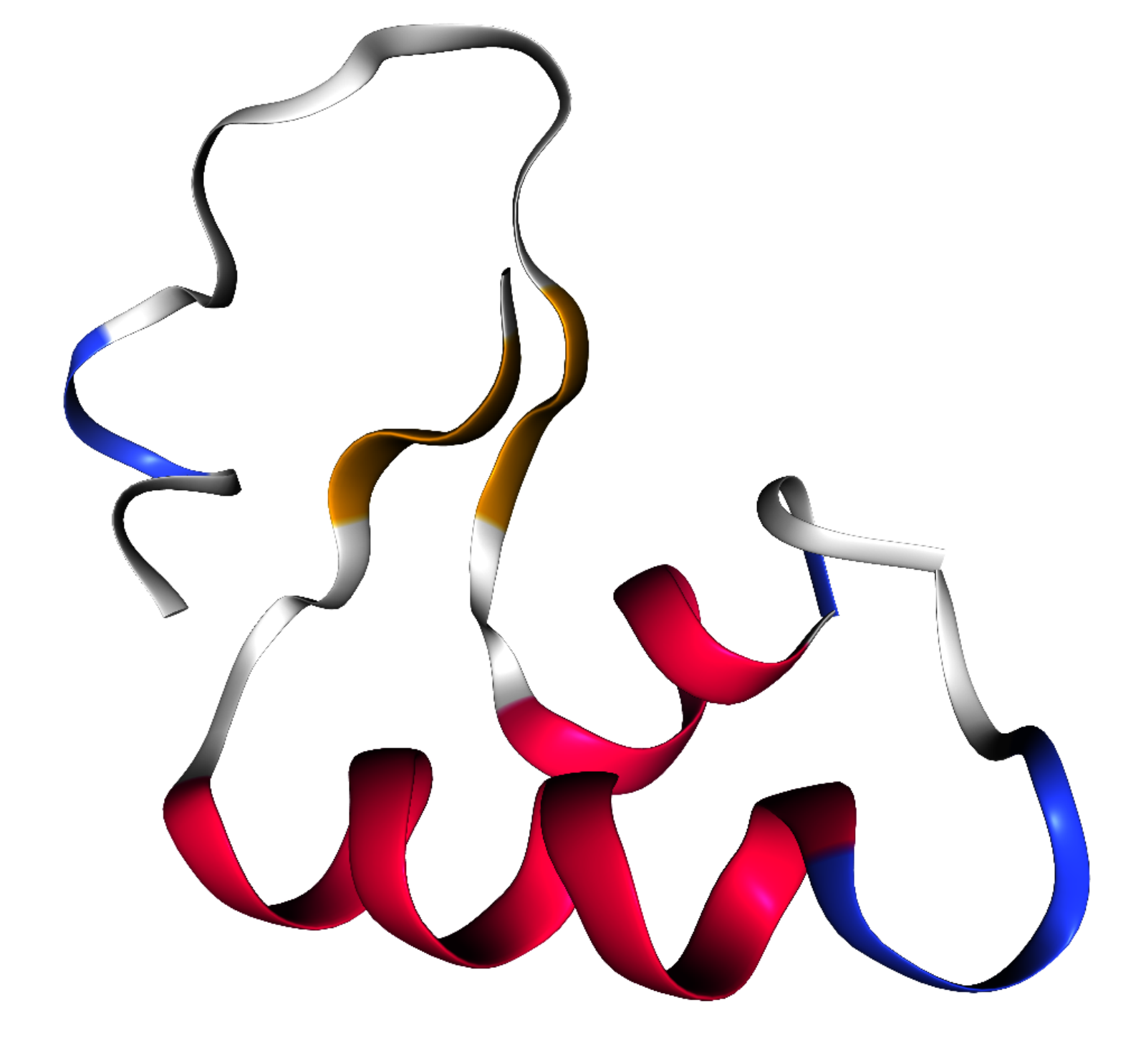}%
    }\hspace{25ex}
    \subfloat[
        \label{fig:crambinNative}
        Model of crambin
    ]{%
        \includegraphics[height=3.5cm]{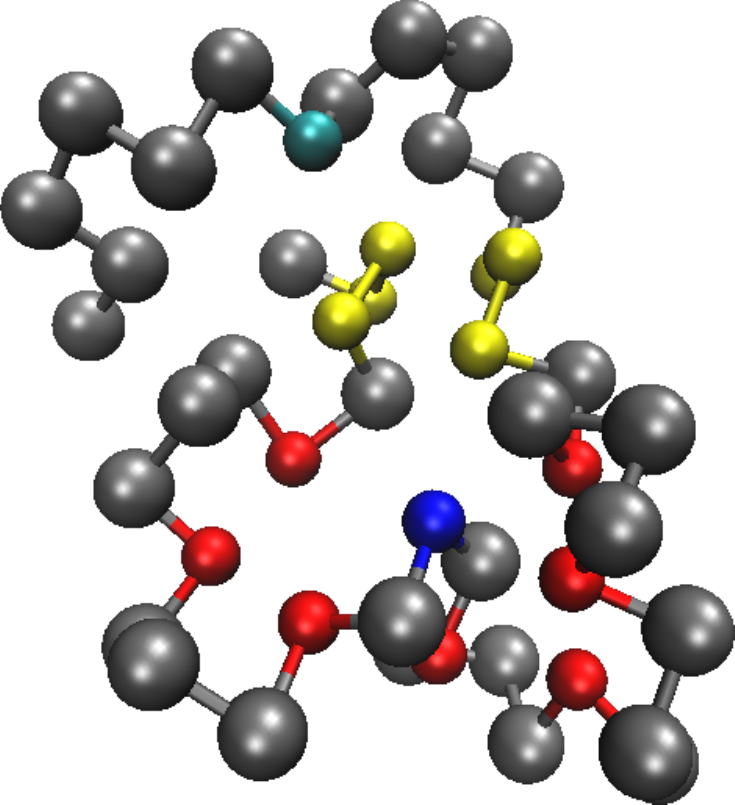}%
    }
    \caption{The model crambin system. Fig.~\ref{fig:crambinPDB} is a cartoon
        representation of the structure of the crystallized protein.
        Fig.~\ref{fig:crambinNative} is the fully-folded minimum entropy state
        of the $46$-bead, ten-bond model. In both figures, the beads
    participating in nonlocal bonds in the $\alpha$-helices are in red, the
$\beta$-sheets are in yellow, and the disulfide bridges are in blue and cyan for
beads $16$ and $40$, respectively.}
    \label{fig:crambinModel}
\end{figure*}

Crambin is a naturally occurring small protein found in the cabbage seeds of
\textit{Crambe abyssinica} consisting of $46$ residues. Because of its ability
to crystallize easily, an unusual characteristic for a protein, the structure of
crambin has been studied using x-ray
crystallography\cite{Schmidt:2011,Jelsch:2000,Teeter:1979}. Despite its small
size, crambin has various secondary structures: two $\alpha$-helices, one
anti-parallel $\beta$-sheet, and three disulfide bridges. In the first
$\alpha$-helix, the bonds span residues $6$ to $18$; in the second helix, they
span residues $22$ to $31$. In the $\beta$-sheet, bonds occur between residues
$[2, 34]$ and $[3, 33]$, and the disulfide bridges occur between residues $[3,
40]$, $[4, 32]$, and $[16, 26]$, as given in the Protein Data Bank. Since
including many nonlocal bonds in the model is computationally expensive to
simulate, we restrict the number of connections in the two $\alpha$-helices of
crambin. Following earlier work\cite{Bayat:2012/245103}, interactions are
allowed between monomers of indices $i = 2 + 4k$ and $j = i + 4n$, where $k$ is
an arbitrary positive integer and $n = 1, 4, 5, \dots$. As a result, crambin's
first $\alpha$-helix is defined by the bonds $[6, 10]$, $[10, 14],$ and $[14,
18]$, and its second helix is defined by the bonds $[22, 26]$, and $[26, 33]$.
Overall, the nonlocal bonds are assumed to occur between beads $[2, 34]$, $[3,
33]$, $[3, 40]$, $[4, 32]$, $[6, 10]$, $[10, 14]$, $[14, 18]$, $[16, 26]$, $[22,
26]$, and $[26, 30]$. All nonlocal bonds are treated as energetically
equivalent, so $\epsilon_{ij} = -\epsilon$, expressed in energy units of $k_B T
\equiv \beta^{-1}$. In Fig.~\ref{fig:crambinModel}, the native state of the
crambin protein from the Protein Data Bank and the fully bonded structure of the
coarse-grained model are shown.

A recent numerical study of the dynamics of the crambin
protein\cite{Colberg:2022} predicted a simple two-step mechanistic folding
pathway in the model protein: Helical regions of crambin formed first with no
clear preference of order, followed by the passage with near-unit probability
through a helical bottleneck state. In the second step, distant bonds linking
helix regions to one another lead to a penultimate state with all bonds present
except those connecting the most distant edges of the chain. In addition, it was
shown\cite{Colberg:2022} that when a diffusive Markov state model is appropriate
for the long-time dynamics of the chain, the effect of different choices of the
nonlocal bond energies $\epsilon_{ij}$ on dynamical properties, such as the mean
first passage time from fully extended configurations to the native structure of
crambin, may be analyzed to determine optimal state energies for this property.
For crambin, it was found that selecting bond energies that decrease the
probability of forming early-stage long-range bonds reduces the folding
probability pathways through the restrictive bottleneck state and doubles the
folding rate relative to the equal bond energy model. In this mechanistic study,
which assumed diffusive dynamics, the overall timescale of the folding process
was written in a scaled time unit, $\ell^2/D$, where $D$ is the self-diffusion
coefficient of the nonlocal bonding distances.

In the present work, transitions among macrostates of a single equal bond energy
model of a crambin molecule in solution are observed in explicit simulations.
Their rates are compared to theoretical predictions based on the diffusive
Markov state model.

\subsection{Thermodynamics} \label{ssec:thermo}

In the coarse-grained model, the configurational space of the protein-like chain
is partitioned into a set of macrostates, where each macrostate is defined by
the bonding pattern or set of satisfied distance constraints. According to the
step-potential for bonding contacts, Eq.~(\ref{eq:attractivestep}), each
nonlocal bond constraint is satisfied or not, depending on whether the pair of
nonlocal beads involved are separated by a distance less than $r_c$. Let the set
of positions of the $N$ monomers be given by the set $\vect{R} = (r_1, \dots,
r_N)$. Since the potential is a square well, the point at which a bond forms or
breaks is well-defined, and each nonlocal bond, $k$, can take on a binary value
$c_k$ of value 0 or 1. If $c_k = 0$, the bond is turned off since $x_k > r_c$,
where $x_k$ is the distance between the nonlocal beads that form bond $k$, and
if $c_k = 1$, the bond is turned on since $x_k < r_c$. Therefore, the set of
nonlocal bonds, $n_b$, for a particular macrostate $c$ out of a total of $n_s =
2^{n_b}$ macrostates can be represented as a binary string:
\begin{equation} \label{eq:binaryString}
    c = c_1, \dots, c_{n_b}.
\end{equation}

Defining an indicator function,
\begin{equation*}
    \mathbbm{1}_c\left(\vect{R}\right) =
    \begin{cases}
        1 & \text{if all constraints for $c$ are satisfied,} \\
        0 & \text{otherwise,}
    \end{cases}
\end{equation*}
the non-ideal configurational entropy of $c$, a dimensionless value, is given by
\begin{equation} \label{eq:entropy}
    S_c = \ln \left(\frac{1}{V^N} \int \mathbbm{1}_c \left(\vect{R}\right)
    d\vect{R} \right).
\end{equation}
The integral over the set $\vect{R}$ can be viewed as the volume of the subspace
of all possible configurations that satisfy the geometric constraints imposed by
the square-well potential and the bonding interactions. As is evident from
Eq.~\eqref{eq:entropy}, $S_c$ is a geometric quantity that depends only on the
distance constraints between the beads; hence, models that differ in the
specified state energies but have the same bonding interaction distances have
the same density of states.

The equilibrium fraction or probability of configuration of $c$, $n_c$, is given
by the canonical average of the indicator function,
\begin{equation*}
    n_c = \langle \mathbbm{1}_c \rangle = \frac{e^{-\epsilon_c}
    e^{S_c}}{\sum_{\alpha=1}^{n_s} e^{- \epsilon_{\alpha}} e^{S_{\alpha}}}
    = \frac{e^{-F_c}}{\sum_{\alpha=1}^{n_s} e^{-F_{\alpha}}},
\end{equation*}
where $F_c$ is the dimensionless free energy of $c$. Since the configurational
entropy is independent of the interaction energy, once $S_c$ has been
determined, the probability of configuration $c$ for any choice of interaction
energies $\{ \epsilon_c | c\}$ (in units of $k_BT$) for the set of states can be
evaluated.

\subsection{The Markov state model of dynamics} \label{ssec:msm}

We assume that each macrostate is long-lived and experiences infrequent but
rapid transitions to other macrostates that differ from it by a single bond. The
set of average non-equilibrium populations, $\vect{n}(t) = \bigl(n_1(t), \dots,
n_{n_s}(t)\bigr)$, defined with respect to an initial distribution,
$P(\vect{R},0)$,
\begin{align*}
    n_i(t) = \int d\vect{R} \, P(\vect{R},0) \mathbbm{1}_i (t),
\end{align*}
evolves according to the Markov state model\cite{Schofield:2014/095101}:
\begin{equation} \label{eq:msm}
    \frac{d\vect{n}(t)}{dt} = \vect{K} \cdot \vect{n}(t),
\end{equation}
where $\vect{K}$ is the transition rate matrix. If state $i$ is the initial
macrostate, with one less bond than state $j$, the final macrostate, the
off-diagonal elements $K_{ji}$ describe a transition between $i$ and $j$ in
which an additional bond is formed when the distance $r$ separating the beads
reaches $r_c$. When this bond is formed, the energy is lowered by $\epsilon$.
Each matrix element is dependent on the mean equilibrium first passage times,
$\tau_{(ij)}^-$ and $\tau_{(ij)}^+$. The inner mean first passage time,
$\tau_{(ij)}^-$, is the average time required for two nonlocal beads in state
$j$ to diffuse to $r_c$ to reach state $i$. In state $j$, $r$ is distributed
according to the conditional equilibrium probability density $\rho_{(ij)}^{-}(r)
= n_j^{-1} H(r_c-r)\rho_i (r)$, where $n_j = \int dr H(r_c-r)\rho_i (r)$ is the
equilibrium population of state $j$, $H(r_c-r)$ is a Heaviside function, and
$\rho_i (r)$ is the marginal equilibrium density for the distance of the bonding
beads when all the bonding constraint conditions in state $i$ are satisfied.
Likewise, the outer first passage time, $\tau_{(ij)}^+$, is the average time for
nonlocal beads diffusing from a distance $r > r_c$, distributed according to
$\rho_{(ij)}^+(r) = n_i^{-1} H(r-r_c)\rho_i (r)$, to reach $r_c$, thereby
yielding a transition from state $i$ to $j$. By examination of the spectrum of
the operator governing the diffusive motion of the system, it can be shown that
the inverse of $K_{ji}$ is\cite{Schofield:2014/095101}
\begin{align} \label{eq:invKji}
    K_{ji}^{-1} &= e^{-\epsilon} e^{\Delta S} \tau_{(ij)}^- +
    \tau_{(ij)}^+ \nonumber \\
    &= \frac{n_i}{n_j} \tau_{(ij)}^- + \tau_{(ij)}^+,
\end{align}
where $\Delta S = S_i - S_j$, and $S_i$ and $S_j$ are the configurational
entropies of $i$ and $j$, respectively. This difference is generally positive
for bond-formation events. The elements of the transition matrix obey detailed
balance,
\begin{equation} \label{eq:detailedbalance}
    K_{ji} n_i = K_{ij} n_j,
\end{equation}
and the probability is conserved such that $\Sigma_j K_{ji} = 0$. It follows
that $K_{ij} = K_{\text{eq}}^{-1} K_{ji}$, where $K_{\text{eq}} = n_j/n_i$ is
the equilibrium constant for the reversible reaction $i \leftrightharpoons j$.
The mean first passage times for the $i$ to $j$ transition can be calculated
using quadratures\cite{Schofield:2014/095101},
\begin{align} \label{eq:mfpt+}
    \tau^{+}_{(ij)} &= \frac{1}{D_{(ij)} }\int_{r_c}^{r_{\text{max}}} \frac{(1 -
    C^{+}_{(ij)}(r))^2}{\rho^{+}_{(ij)}(r)} dr =
    \frac{\overline{\tau}^{+}_{(ij)}}{D_{(ij)}}, \\
    \tau^{-}_{(ij)} &= \frac{1}{D_{(ij)}} \int_{r_{\text{min}}}^{r_c} \frac{
    C^{-}_{(ij)}(r)^2}{\rho^{-}_{(ij)}(r)} dr =
    \frac{\overline{\tau}^{-}_{(ij)}}{D_{(ij)}},
    \label{eq:mfpt-}
\end{align}
where $r_{\text{min}}$ and $r_{\text{max}}$ are the minimum and maximum allowed
values of the bonding distance that separates two nonlocally bonded beads, and
$\overline{\tau}^{\pm}_{(ij)}$ defines a mean squared diffusion distance with
units $\ell^2$. In Eqs.~\eqref{eq:mfpt+} and \eqref{eq:mfpt-}, $D_{(ij)}$ is
the diffusion coefficient characterizing the diffusive evolution of the relative
bond distance in the absence of the bonding potential, and the cumulative
distributions are defined as\cite{Schofield:2014/095101}
\begin{align}
    C_{(ij)}^{-}(r) &= \int_{r_{\text{min}}}^r \rho^{-}_{(ij)}(x) \, dx,
    \nonumber \\
    C_{(ij)}^{+}(r) &= \int_{r_c}^{r} \,\rho^{+}_{(ij)}(x) \, dx. \nonumber
\end{align}
In practice, the diffusion coefficients $D_{(ij)}$, which depend on both
solvent-bead and local bead-bead interactions, are similar for most pairs of
configurations $(ij)$ in the model (see Table~\ref{table:layers}). However, the
diffusion coefficients of individual beads depend on the viscosity of the
solvent environment in which the protein-like chain is embedded, and this
viscosity can differ by many orders of magnitude.

As discussed in the Appendix, the validity of Eq.~\eqref{eq:invKji} depends on a
number of assumptions: In particular, there must be a separation between the
timescale on which contact bonds are formed and broken, leading to a change in
the macrostate, and the timescale of the local equilibration of each macrostate.
Additionally, the force exerted by the solvent-bead and bead-bead interactions
on each bead must be rapid on the timescale of bead motion, and the friction
must be strong enough to lead to overdamped bead motion in which velocity
correlations decay to zero on a timescale that is shorter than the typical
timescale of changes in the positions of the beads.

\subsection{The layer method} \label{ssec:layer}

Colberg and Schofield\cite{Colberg:2022} developed a layer method in which an
adaptive dynamical Monte Carlo scheme was used to obtain an estimate of $S_c$ in
Eq.~\eqref{eq:entropy} for each intermediate state as well as to compute
numerical estimates of the inner and outer first passage times. The procedure
consists of successively evaluating in parallel the entropy difference and first
passage times between states that differ by the formation of a single contact
using adaptive sampling that combines a Wang-Landau algorithm with a rigorous
statistical convergence criterion to establish confidence intervals for all
quantities. When a model protein has many possible nonlocal bonds, the number of
possible macrostates, $n_s$, which grows exponentially, is prohibitively large.
The large number of states negatively impacts the statistical resolution of
adaptive sampling since the set of standard errors of observed quantities
decreases\cite{Colberg:2022} at a rate proportional to $n_s^{-1/2}$. The layer
method improves the efficiency of the calculation of the density of states by
evaluating the differences in entropy of many transitions in parallel while
restricting the active set of states to $n_s = 2$.

During each of the sampling iterations for a particular pair of connected
macrostates, the distance between the bonding beads in the active bonds is
recorded and used to construct analytical fits of the marginal equilibrium
probability densities $\rho_{(ij)}^{\pm} (r)$ and their cumulative distribution,
based on a maximum likelihood estimate of the logarithm of the density using
splines\cite{Schofield:2017}. The mean squared diffusion distances defined in
Eqs.~(\ref{eq:mfpt+}) and (\ref{eq:mfpt-}) are then obtained by numerical
integration of the analytical fits.

\subsection{The diffusion coefficients} \label{ssec:diffcoeff}

In a diffusive Markov state model, all elements of the transition rate matrix
$\vect{K}$ are proportional to the diffusion coefficient characterizing the
evolution of the bonding distance between the relevant nonlocal bond-forming
beads. The values of these diffusion coefficients, and hence the timescales of
the changes in the macrostates, depend strongly on the nature of the solvent in
which the protein-like chain is immersed.

In general, for single mesoscale-sized monomers of large size $R$ and mass $M$,
the self-diffusion coefficient, defined in terms of the time integral of the
equilibrium velocity autocorrelation function, can be written as the sum of two
components\cite{Hynes:1979},
\begin{align} \label{eq:diffcoeff}
    D &= \frac{1}{3} \int_0^\infty d\tau \; \langle \vect{V}(\tau) \cdot
    \vect{V} \rangle \nonumber \\
    &= D_0 + D_h,
\end{align}
where $\vect{V}$ is the velocity of the monomer and $D_0$ is a bare diffusion
coefficient arising from uncorrelated collisions between the monomer and the
solvent. In Eq.~\eqref{eq:diffcoeff}, $D_h$ is a hydrodynamic contribution that
accounts for the dynamic correlations induced by the motion of the monomer in
the fluid, which is also responsible for the algebraic decay of the velocity
autocorrelation function (VACF) on hydrodynamic
timescales\cite{Alder:1970,Schofield:1992}. For a spherical monomer in a
hard-sphere fluid, the bare diffusion coefficient, $D_0$, arising from the rapid
loss of velocity correlations due to solvent collisions on molecular timescales,
is approximately given by the Enskog result\cite{Hynes:1979,Stell:1982},
\begin{align} \label{eq:bareDiffusion}
    D_0 &= \frac{k_B T}{\chi_0}, \nonumber \\
    \chi_0 &= \frac{8 \rho}{3} R^2 g(R) \sqrt{2 \mu_m \pi k_B T},
\end{align}
where $k_B$ is Boltzmann's constant, $T$ is the temperature of the system,
$\rho$ is the number density of the solvent, $m$ is the mass of a solvent
particle, $\mu_m = Mm / (M + m)$ is the reduced mass, and $g(R)$ is the radial
distribution function at contact\cite{Silva:1998}. Asymptotically, when the
size of the monomer $R$ is much larger than the correlation length $\xi$ induced
by the solvent-monomer interaction potential (i.e. the boundary
layer)\cite{Schofield:1992}, an approximate form of the hydrodynamic
contribution for an unbounded system is given by the Stokes' result in the slip
limit\cite{Bocquet:1994a,Bocquet:1994},
\begin{align}
    D_h &= \frac{k_B T}{\chi_h}, \nonumber \\
    \chi_h &= 4 \pi \eta_d R, \label{eq:StokesDiffusion}
\end{align}
where $\eta_d$ is the dynamic viscosity and $R$ is the hydrodynamic radius. For
large Brownian particles, such as in the case of a large sphere suspended in a
hard-sphere fluid, $D_{h} \gg D_0$, since $D_0$ and $D_{h}$ scale differently
with $R$, the hydrodynamic size of the monomer. The clear separation between
molecular scale and hydrodynamic contributions to diffusion breaks down as the
size of the monomer approaches the thickness of the boundary layer, $R \sim
\xi$. Furthermore, for finite systems, such as those in computer simulations,
important correction terms to the hydrodynamic friction $\chi_h$ arise from the
long-ranged nature of the hydrodynamic interactions\cite{Celebi:2021}.

The derivation of microscopic expressions for the diffusion coefficient
$D_{(ij)}$ for the relative bond distance in the protein-like chain is
complicated by the contribution of the local bead-bead interactions to the
non-hydrodynamic friction, $\chi_0$. The collisions of a monomer with its local
nearest and next-nearest neighbors can occur on a timescale that is not clearly
separated from the timescale of collisions between the monomer and the solvent,
resulting in negative velocity correlations and a reduction in the
self-diffusion coefficient as observed in dense fluids in which caging is
prominent. The local interactions typically result in a $D_{(ij)}$ that differs
significantly from the case in which monomers are not connected in a chain, in
which case $D_{(ij)} = 2D_0$. In the absence of theoretical results for such
systems, the bond distance diffusion coefficients must be evaluated by direct
numerical simulation.

\section{MODELS OF THE SOLVENT BATH} \label{sec:solventbath}

The modeling of a solvent bath faces the same challenges as simulating the chain
itself. Atomistic calculations are computationally expensive, and the
microscopic timescales of the monomer motions are orders of magnitude shorter
than the timescale on which conformational changes occur in biological
systems\cite{Gompper:2009}. Coarse-graining can recreate the behavior of an
atomistic system at timescales that more closely resemble biological systems.
Some coarse-grained approaches, such as a continuum model based on the Stokes
equation\cite{Gompper:2009}, are too simplistic since they implicitly omit
important physical effects such as thermal fluctuations, which play an important
role in the motion of the solvent and small solutes. We will use three models of
the solvent of decreasing complexity that progressively simplify physical
phenomena: A hard-sphere model, a multi-particle collision model that neglects
fluid structure, and, finally, a penetrating solvent model in which an effective
fluid is simulated through stochastic collisions that periodically modify the
bead velocities while neglecting hydrodynamic interactions.

\subsection{The hard-sphere solvent model} \label{ssec:hardsphmodel}

In a hard-sphere model, the treatment of the solvent particles is similar to
that of the beads in a protein-like chain: Each particle is represented by a
hard-sphere, which travels in a straight line until it experiences an elastic
collision due to an encounter with another particle. As a result, hydrodynamic
flow and solvent-solvent interactions are present in the hard-sphere model.
Since the hard-sphere solvent particles occupy a non-zero volume, unlike point
particles, the physical configurations of the solvent require non-overlapping
solvent spheres.

Theoretical predictions for the self-diffusion coefficient of a large and
massive monomer in a hard-sphere fluid can be obtained from kinetic
theory\cite{Bocquet:1994a}. Accurate expressions\cite{Smith:2002} of the radial
distribution function at contact, $g(R)$, can be used in the Enskog expression
in Eq.~(\ref{eq:bareDiffusion}) to evaluate the bare diffusion coefficient
$D_0$, and the hydrodynamic contribution to the self-diffusion coefficient of
the monomer, $D_h$, can be estimated using Enskog results for the dynamic
viscosity\cite{Stell:1982}. The predicted forms of the self-diffusion
coefficient in a hard-sphere fluid are useful to determine the physical
conditions, such as the particle size and number density, which are required for
modeling fluids with specified physical characteristics.

\subsection{The multi-particle collision dynamics solvent model}

MPCD, also known as stochastic rotation dynamics (SRD), is a solvent model
developed by Malevanets and Kapral in which the solvent is coarse-grained to
eliminate the calculation of solvent-solvent
interactions\cite{Malevanets:1999,Malevanets:2000,Kapral:2008,Gompper:2009}.
Therefore, intermolecular interaction potentials between solvent particles and
structural ordering in the fluid cannot occur\cite{Schofield:2012}. MPCD is
carried out in two alternating steps: a solvent streaming step, in which solvent
particles move in the presence of the solute, and a stochastic solvent collision
step. In this model, the solvent's mass, momentum, and energy are conserved
locally, resulting in dynamics consistent with hydrodynamic flow. Due to the
conservation of energy, MPCD can be used for simulating microcanonical
ensembles\cite{Gompper:2009}. Both hydrodynamic flow and thermal fluctuations
are included in this model, but the collisions between the solvent particles are
carried out at specific steps rather than continuously throughout the
simulation.

In the hard-sphere MPCD model, the solvent consists of a large number $N_s$ of
particles, each with mass $m$\cite{Kapral:2008,Gompper:2009}. These particles
move independently and interact with the monomers of the solvated chain via
elastic collisions. Solvent-solvent interactions occur via collision steps at
discrete time intervals, $\Delta t$, which, unlike the time intervals in
molecular dynamics, do not need to be small\cite{Gompper:2009}. Let the volume
occupied by the system of solvent particles be divided into a grid made up of
$N_{x}$ cubic cells, each with side length $a$\cite{Kapral:2008,Gompper:2009}.
Within each cell $\phi$, there are $N_\phi$ solvent particles. Each particle can
only interact with its cell mates and not with the particles in neighboring
cells. The particles can also cross from one cell to another; thus, the quantity
$N_\phi(t)$ is dynamic. The cells have a center of mass velocity,
$\vect{V}_\phi(t)$, where
\begin{equation}
    \vect{V}_\phi(t) = \frac{1}{N_\phi(t)} \sum_{j=1}^{N_\phi(t)}
    \vect{v}_{j}(t),
\end{equation}
where each particle $j$ in the set of $N_\phi (t)$ particles is positioned in
cell $\phi$ at the rotation time.

The post-collision velocity at collision time $t$, $\vect{v}'_i(t)$, of particle
$i$ located in a cell $\phi$ with center of mass velocity $\vect{V}_\phi(t)$,
is given by
\begin{equation} \label{eq:mpcdfinalvel}
    \vect{v}'_i(t) = \vect{V}_\phi(t) + \hat{\vect{\omega}}_\phi
    \cdot \left(\vect{v}_i(t) - \vect{V}_\phi(t)\right),
\end{equation}
where $\hat{\vect{\omega}}_\phi$ is a rotation operator assigned to cell $\phi$
chosen randomly from a set of rotation operators,
$\Omega$\cite{Kapral:2008,Gompper:2009}. Therefore, the relative particle
velocities within a given cell undergo the same rotation, while the rotation
matrices $\hat{\vect{\omega}}_\phi$ for each cell differ. There are several ways
in which a rotation can be carried out. Here, the particle velocities in each
cell are rotated by an angle $\alpha$ about a randomly chosen direction
$\hat{\vect{u}}$. Therefore, the post-collision velocity of $i$ in a cell
$\phi$ can be rewritten as
\begin{align} \label{eq:mpcd-finalvel2}
    \vect{v}'_i(t) &= \vect{V}_{\phi}(t) + \hat{\vect{u}}
    \hat{\vect{u}} \cdot \left(\vect{v}_i(t) - \vect{V}_{\phi}(t)\right)
    \nonumber \\
    &+ \left(\vect{I} - \hat{\vect{u}} \hat{\vect{u}}\right) \cdot
    \left(\vect{v}_i(t) - \vect{V}_{c}(t)\right) \cos{\alpha} \nonumber \\
    &- \hat{\vect{u}} \times \left(\vect{v}_i(t) - \vect{V}_{\phi}(t)\right)
    \sin{\alpha}.
\end{align}
Finally, prior to the collision step, the entire cell grid is shifted by a
uniform random displacement vector to preserve Galilean
invariance\cite{Ihle:2001,Ihle:2003,Gompper:2009}.

For an MPCD fluid in which fluid particles interact with the monomer beads by a
hard-sphere potential, the bare diffusion coefficient $D_0$ for a single monomer
is given by Eq.~(\ref{eq:bareDiffusion}) with $g(R)=1$. The hydrodynamic
contribution $D_h$ can be approximated using the explicit form for the dynamic
viscosity, $\eta_d = m \rho (\eta_k + \eta_c)$, where $\eta_k$ and $\eta_c$ are
the kinetic and collisional contributions to the viscosity\cite{Gompper:2009}
given by
\begin{align} \label{eq:kscontrib}
    \eta_k &= \left[\frac{5N_c}{\left(N_c-1+e^{-N_c}\right)
    \left[2-\text{cos}(\alpha)-\text{cos}(2\alpha)\right]} - 1\right]
    \frac{k_B T \Delta t}{2m} \nonumber \\
    \eta_c &= \left[\frac{\left(N_c-1+e^{-N_c}\right)}{18N_c} \left[1 -
    \text{cos}(\alpha)\right]\right] \frac{a^2}{\Delta t},
\end{align}
and $N_c = \rho a^3$ is the average number of particles in one cubic cell.

\subsection{The penetrating solvent model} \label{ssec:pensolmodel}

In the penetrating solvent model, the solvent interactions are implicitly
included by creating fictitious solvent-bead
collisions\cite{Gompper:2009,Schofield:2012}. The interactions between solvent
particles and monomers occur through rotations of the monomer velocities
relative to an effective center of mass velocity,
\begin{equation} \label{eq:pencom}
    \vect{V}_\phi(t) = \frac{M}{M_t} \vect{v}_i(t) + \frac{N_s m_s}{M_t}
    \vect{v}_s(t),
\end{equation}
where $N_s$ is randomly chosen from a Poisson distribution with a mean of $N_c =
\rho V_{\text{cell}} = \rho a^3$. In Eq.~\eqref{eq:pencom}, $M_t = M + N_s m$,
and $\vect{v}_s$ is chosen from a normal distribution with a zero mean and a
variance of $k_B T / \left(N_s m_s\right)$, where $T$ is the system's
temperature. Post-collision, the velocity of the bead is given by
Eq.~\eqref{eq:mpcd-finalvel2}.

Unlike the previous two models, there is no hydrodynamic contribution, $D_h$, to
the diffusion coefficient of the penetrating solvent model as there is no fluid
velocity field. Rather, the value of $D = D_0$ arises purely from the
uncorrelated collisions between the fictitious solvent and bead particles. By
explicitly calculating the VACF and its integral, one
finds\cite{Schofield:2012}:
\begin{equation} \label{eq:expdiff}
    D = \frac{k_B T}{M} \Delta t \left(\frac{2 - \gamma}{2 \gamma}\right),
\end{equation}
where $\gamma$ is given by
\begin{equation} \label{eq:gamma}
    \gamma = \frac{2N_c}{3(1 + \mu)} M(1, 2 + \mu, -N_c),
\end{equation}
and $M(1, 2 + \mu, -N_c)$ is Kummer's function of the first
kind\cite{Abramowitz:1965}. Here, $\mu = M/m$ is the ratio of the bead and
solvent masses.

\section{PARAMETERIZATION OF SIMULATIONS} \label{sec:simparam}

\begin{figure}[h!]
  \centering
  \includegraphics[width=\linewidth]{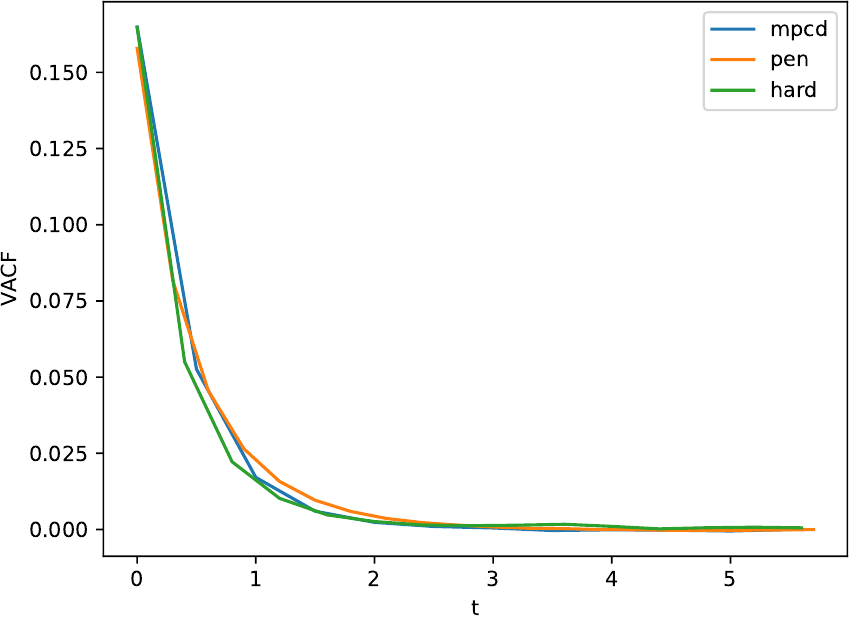}
  \caption{Unnormalized velocity autocorrelation curves $\langle \bm{V}(t) \cdot
  \bm{V} \rangle/3$ vs. time (in simulation units $\tau$) for a Brownian
  particle diffusing in a bath.}
  \label{fig:vacf}
\end{figure}

To compare the population equilibration rate for model proteins immersed in the
hard-sphere, MPCD, and penetrating solvent models, the simulation parameters
must be chosen such that monomer dynamics are similar in the three models. To
select these parameters, a massive (Brownian) bead in equilibrium with the
solvent was simulated in the respective solvent models. Each parameter was
designed to yield closely matched velocity autocorrelation functions and
diffusion coefficients. In addition, the mean-squared displacements (MSD) of the
Brownian bead, $\left<|\vect{X}(t)-\vect{X}(0)|^2\right>$, were monitored. All
particles were confined to a cubical box of side length $L$ with periodic
boundary conditions in all directions. For the penetrating and MPCD models, $L =
10 \ell$, where $\ell$ is the simulation unit of length, whereas for the
hard-sphere model, $L = 9 \ell$. The slightly smaller box size for the latter
model was chosen to reduce the number of hard-sphere solvent particles needed
due to the high computational cost of simulating the monomer-solvent and
solvent-solvent collisions. In the penetrating and MPCD models, the number
density was $\rho = 8.87 \ell^{-3}$, and in the hard-sphere model, the number
density was $\rho = 10 \ell^{-3}$. For all models, the mass $M$ of the Brownian
particle was set to be six times the mass of the solvent particles $m$, taken to
be the unit of mass since an amino acid weighs, on average, six times more than
a water molecule. The rotation time $\Delta t$ was taken to be $0.25 \, \tau$
for the penetrating solvent model, while for the MPCD model, it was $\Delta t =
0.5 \, \tau$, where $\tau = \sqrt{m\ell^2/(k_B T)}$ is the simulation time unit.
In both cases, the rotation angle was $\alpha = 90^{\circ}$. The radius of the
Brownian particle was $0.5 \ell$ for the penetrating and MPCD models, whereas
for the hard-sphere model, it was $0.4 \ell$. In the MPCD solvent, the cell
length $a = \ell$. The radius of each hard-sphere solvent particle was set to
$0.1 \ell$, since water molecules are approximately three to four times smaller
than an amino acid. As a result of the choice of these parameters, the mean free
path for the MPCD model was $\lambda = \Delta t / \tau \cdot \sqrt{k_B T / m} =
0.5 \ell$. For these parameters, $\chi_h/ \chi_0 \sim b \ell/(Rg(R)) [(5\Delta
t)/(4\tau) + \tau/(9\Delta t)] \sim 4.77$, where $b = 18\pi/\sqrt{32 \pi}$, so
the hydrodynamic contribution $D_h$ to the diffusion coefficient of a single
bead is expected to be smaller than the direct interaction contribution, $D_0$.
For larger values of $R/\ell$, the hydrodynamic contribution $D_h$ dominates
$D_0$. Using these parameters, the resulting unnormalized VACF curves from all
three solvent models, the integral of which defines the self-diffusion
coefficient according to Eq.~(\ref{eq:diffcoeff}), were found to be nearly
identical when overlaid, as can be seen in Fig.~\ref{fig:vacf}.

The parameters prescribed earlier were used in equilibrium simulations to obtain
diffusion coefficients for all three solvent models. For the Brownian system,
the diffusion coefficient was evaluated in two ways: by computing the running
integral of the VACF, and by a least squares polynomial fit of the short, linear
portion of the mean squared displacement (MSD). The two expressions for the
diffusion coefficient are related by noting that for $t \gg \tau$, the MSD
satisfies
\begin{align} \label{eq:MSD2}
   \left<|\vect{X}(t) - \vect{X}(0)|^2\right> &= 2t \int_0^t d\tau \left(1 -
   \tau / t\right) \left<\vect{V}(\tau) \cdot \vect{V}(0)\right> \nonumber \\
   &\approx 6 D(t) \, t,
\end{align}
where
\begin{align} \label{eq:vacfIntegral}
    D(t) &= \frac{1}{3} \int_0^t d\tau \left\langle \vect{V}(\tau) \cdot
    \vect{V} \right\rangle.
\end{align}
The diffusion coefficient estimates for each solvent model are recorded in
Table~\ref{table:browniandiffs}.

\begin{table}
    \begin{tabular}{l l l}
        \toprule
        Solvent model & $D_{\text{MSD}}$ & $D_{\text{VACF}}$ \\
        \midrule
        Penetrating & 0.080 & 0.085 \\
        MPCD & 0.074 & 0.079 \\
        Hard-sphere & 0.073 & 0.080 \\
        \bottomrule
    \end{tabular}
    \caption{Diffusion coefficients in units of $\ell^2/\tau$ of the Brownian
    system immersed in the penetrating, MPCD, and hard-sphere solvent models.}
    \label{table:browniandiffs}
\end{table}

The typical value of the diffusion coefficient was found to be $D = 0.08 \,
\ell^2/\tau$, in simulation units. Appropriate length, mass, and energy scales
must be identified to compare this diffusion coefficient value for a Brownian
monomer to experimental values. Theoretical estimates of the diffusion
coefficient in protein systems typically rely on the Stokes-Einstein law (i.e.
Eq.~(\ref{eq:StokesDiffusion})), which requires the input of the effective
spherical hydrodynamic radius and the viscosity of the solvent\cite{Brune:1993}.
For monomers or groups of atoms {\it within} a protein, neither the choice of
the effective hydrodynamic radius $R$ nor the local viscosity of the monomer in
the fluid is apparent. Experiments generally measure the protein's mass
diffusion center rather than the diffusion coefficient of monomers or groups of
atoms, and typical values of $5 \cdot 10^{-10}$ $m^2/s$ are
found\cite{Walderhaug:2010,Evans:2018}. The length scale $\ell$ is set to the
average length of a covalent bond between two amino
acids\cite{Bayat:2012/245103}, $\ell = 3.84$ \angstrom, and estimating the
average solvent mass $m$ in the cytoplasm to be on the order of $100$ amu, $\tau
= 2.27 \cdot 10^{-12} \, \text{s}$ and $D \approx 5 \cdot 10^{-9} \text{
m}^2/\text{s}$ in standard units. This value of the diffusion coefficient is
roughly ten times larger than estimates of the diffusion coefficient of the
center of mass of large
substrates\cite{Brune:1993,Walderhaug:2010,Schofield:2012,Evans:2018,Tang:2022}.
In dilute solutions\cite{Nemoto:1984}, the center of mass self-diffusion
coefficient of a polymer scales with the number of monomers $N$ as $N^{-0.55}$,
so the monomer diffusion measured in the simulation would correspond to a small
protein of roughly $65$ monomers, roughly the size of the crambin protein. For
the penetrating and MPCD fluid models, the monomer diffusion coefficients can be
reduced by decreasing the rotation time $\Delta t$, hence increasing the
frequency of stochastic rotations, with a proportional increase in the
computational burden of simulating the solvated system. Increasing the number
density of hard-sphere fluid particles similarly leads to a roughly proportional
increase in the bare and hydrodynamic friction, $\chi_0$ and $\chi_h$,
respectively, while increasing the computational burden at least quadratically.

\section{SIMULATED DYNAMICS} \label{sec:simulations}

To verify the diffusive Markov state model of the dynamics for the different
solvents, we consider the dynamics of a single layer in which a single transient
bond is allowed to form and break to allow transitions between the unbonded
state $u$ and the bonded state $b$. Denoting the relative equilibrium population
of the unbonded state as $n_u^{\text{eq}} = n_u/(n_u+n_b)$ and its instantaneous
population as $n_u(t)$, the two-state Markovian dynamics for the population
difference, $\delta n_u(t) = n_u(t) - n_u^{\text{eq}}$, and instantaneous
population, $n_u(t)$, are given by
\begin{align} \label{eq:solofode}
    \delta n_u(t) &= e^{-(K_{bu} + K_{ub}) t} \delta n_u(0) = e^{-k_{\text{r}}t}
    \delta n_u(0), \nonumber \\
    n_u(t) &= n_u^{\text{eq}} + \left(n_u(0) - n_u^{\text{eq}}\right)
    e^{-k_{\text{r}} t},
\end{align}
where the decay rate $k_{\text{r}}$ from an initial non-equilibrium population
difference $\delta n_u(0)$ is $k_r = K_{bu} + K_{ub} = K_{bu} (1 +
K_{\text{eq}}^{-1})$.

\begin{figure}[h!]
  \centering
  \includegraphics[width=\linewidth]{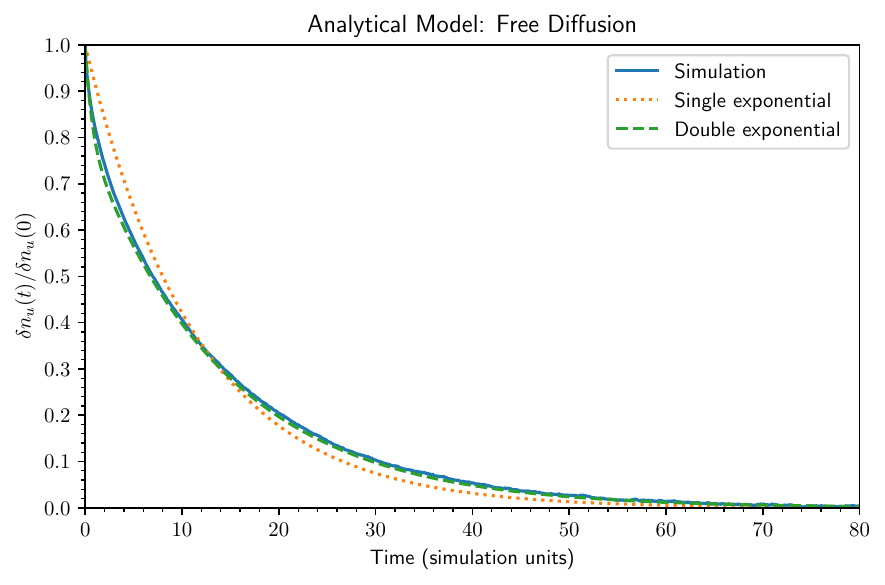}
  \caption{A relaxation profile for an ensemble of $N_e = 10^5$ initially
      unbonded free particles (state $u$) with a mass ratio of $\mu=20$ and
      diffusion coefficient $D = 0.0205 \, \ell^2/\tau$, corresponding to a
      rotation time $\Delta t = 0.005 \, \tau$, in an implicit solvent of number
      density $\rho = 10 \,\ell^{-3}$. A reflecting boundary at $R=2$ confines
      the particles, and the particles experience a discontinuous drop in the
      potential of $\epsilon = k_B T$ at $R=1$. The solid line is the simulation
  result, the dotted lines are analytical predictions for the decay using
  Eq.~(\ref{eq:invKji}), and the dashed lines are a higher-order Pad\'e
  approximation to $\delta n_u(t)/\delta n_u(0)$, as discussed in the Appendix.}
  \label{fig:freeModel}
\end{figure}

Fig.~(\ref{fig:freeModel}) shows a typical relaxation profile of an ensemble of
particles with a mass ratio $M/m = \mu = 20$, initially distributed in an
unbonded equilibrium state for an analytically solvable model in which each
particle moves freely and independently in a penetrating solvent (described in
Sec.~\ref{ssec:pensolmodel}), and experiences a discontinuous drop in potential
energy of $\epsilon = 1$ at a radial distance $R=1$ from the origin inside a
sphere with a reflecting boundary at $R=2$. As is evident in
Fig.~(\ref{fig:freeModel}), although the relaxation profile is expected to
deviate from a single exponential for a diffusive process, most notably at short
times, the overall timescale of relaxation is well-described by a single
exponential of relaxation time $k_r^{-1}$. As discussed in the Appendix, if
desired, the predicted profile can be improved by a higher-order Pad\'e
approximation that includes the initial rate of decay of the populations and its
integrated first moment.

The coarse-grained model of crambin has $n_b = 10$ bonds, which gives $n_s =
1024$ possible macrostates. There are a total of $5120$ possible transitions
and, therefore, individual simulations. While simulating the entire set of
transitions may be feasible in some cases, it is computationally expensive to do
for all but the simplest models with a solvent present. As a result, in this
work, ten representative transitions for crambin were chosen, each from a
different layer, corresponding to a different number and set of permanently
maintained bonds. The position of the active bond in each transition is also
different. The selected transitions include qualitatively different transition
elements: In each higher layer, the initial state was chosen to have different
bonding patterns from the pattern studied in the lower layers (with fewer
nonlocal bond constraints). The layer number, initial and final bit patterns of
the intermediate states in these transitions, the difference in entropy between
the respective states, and the inner and outer first passage times computed
using the layer method as described in Sec.~\ref{ssec:layer} are given in
Table~\ref{table:layers}.

\begin{table}
    \begin{tabular}{l l l l l l l}
        \toprule
        Layer & Initial State & Final State & $\Delta S$ &
        $\overline{\tau}_{(ij)}^-$ & $\overline{\tau}_{(ij)}^+$ & $D_{(ij)}$ \\
        \midrule
        0 & 0000000000 & 0000000001 & 3.68 & 0.0170 & 3.1 & 0.0403 \\
        1 & 1000000000 & 1000010000 & 3.56 & 0.0173 & 2.9 & 0.0400 \\
        2 & 0000100100 & 0000100110 & 3.24 & 0.0172 & 2.0 & 0.0308 \\
        3 & 1000000110 & 1010000110 & 6.97 & 0.0146 & 82 & 0.046 \\
        4 & 0100101100 & 1100101100 & 3.92 & 0.0151 & 3.1 & 0.0394 \\
        5 & 0011110010 & 0011110110 & 5.82 & 0.0138 & 21 & 0.0362 \\
        6 & 0011111010 & 0111111010 & 3.69 & 0.0148 & 2.4 & 0.027 \\
        7 & 0110110111 & 0111110111 & 3.07 & 0.0163 & 1.54 & 0.027 \\
        8 & 1111001111 & 1111101111 & 3.25 & 0.0179 & 2.3 & 0.0358 \\
        9 & 1111110111 & 1111111111 & 3.82 & 0.0160 & 3.1 & 0.027 \\
        \bottomrule
    \end{tabular}
    \caption{The layer number and bit patterns of crambin's initial and final
        states simulated in the presence of a solvent. The bits indicate which
        bonds are formed in the respective states in the $46$ bead protein model
        according to Eq.~(\ref{eq:binaryString}), where the indexing refers to
        the set of ten bonds: $[2, 34]$, $[3, 33]$, $[3, 40]$, $[4, 32]$, $[6,
        10]$, $[10, 14]$, $[14, 18]$, $[16, 26]$, $[22, 26]$, and $[26, 30]$.
        The percent error for $\Delta S$ is less than $5 \%$; for
        $\overline{\tau}_{(ij)}^-$ and $\overline{\tau}_{(ij)}^+$, it is $5 \%$;
        and for $D_{(ij)}$, it ranges from $1 \%$ to $6 \%$. The diffusion
        coefficients $D_{(ij)}$ are the simulation results for the penetrating
    solvent model.}
    \label{table:layers}
\end{table}

\subsection{The bond distance diffusion coefficient} \label{ssec:simdiffcoeff}

To evaluate the diffusion coefficients, $D_{(ij)}$, for the bonding distances
that are critical to identifying the timescale of structural transitions in the
Markov state model, simulations for each of the layers of crambin given in
Table~\ref{table:layers} were run with $\epsilon = 0.0$ for each type of
solvent. To compute $D_{(ij)}$, the positions of the two beads in the active
bond were recorded at regular time intervals, and the MSD was calculated. As
with the Brownian particle, a least squares polynomial fit was performed over a
linear portion of the MSD data. This was done at intermediate timescales $\tau_i
\sim 2 \tau \ll k_r^{-1}$ that were longer than the ballistic timescale but
shorter than the timescale of bond formation. From this fit, the diffusion
coefficient was obtained and used to predict the values of the Markov state
model rates $k_{\text{r}}$ and the equilibrium population $n_u^{\text{eq}}$ for
the chosen transitions.

In the absence of all local bonding interactions that contribute an internal
source of friction, we expect that $D_{(ij)} \approx 2D_0$, where $D_0 \sim 0.08
\, \ell^2/\tau$, is the diffusion coefficient for a single monomer. From the
values in Table~\ref{table:layers}, it is evident that the
configurationally-dependent local interactions produce internal friction that
significantly reduces the bond distance diffusion coefficient, and $D_{(ij)}$
tends to smaller values at higher layers where more permanent bonds restrict the
motion along the bond formation coordinate.

\subsection{Analysis of simulations} \label{ssec:analysis}

To obtain numerical estimates for the $k_{\text{r}}$ and $n_u^{\text{eq}}$ for
transitions between crambin macrostates, simulations were carried out for all
layers in Table~\ref{table:layers} at three values of $\epsilon$: $3.0$, $4.0$,
and $5.0$. These values of $\epsilon$ yielded a good balance of sufficiently
accurate data and fast simulation times.

\begin{figure}[htbp]
  \centering
  \includegraphics[width=\linewidth]{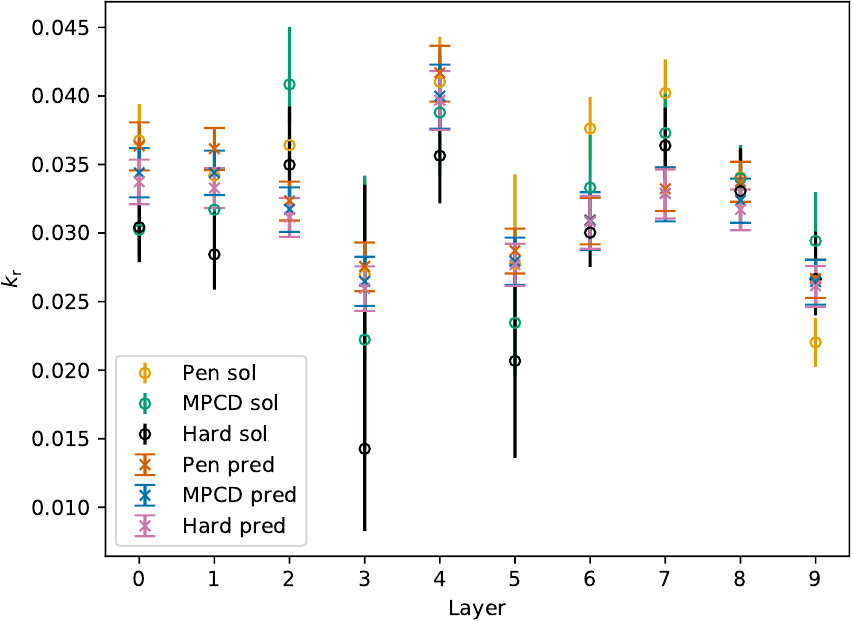}
  \caption{The $k_{\text{r}}$ for crambin for all three solvent models at
  $\epsilon = 3.0$. In the legend, the keywords \textit{pen sol}, \textit{mpcd
  sol}, and \textit{hard sol} correspond to the $k_r$ values from the
  penetrating, MPCD, and hard-sphere solvent models respectively, and
  \textit{pen pred}, \textit{mpcd pred}, and \textit{hard pred} represent the
  rates predicted using the diffusive Markov state model.}
  \label{fig:oldparamkrates}
\end{figure}

\begin{figure}[htbp]
  \centering
  \includegraphics[width=\linewidth]{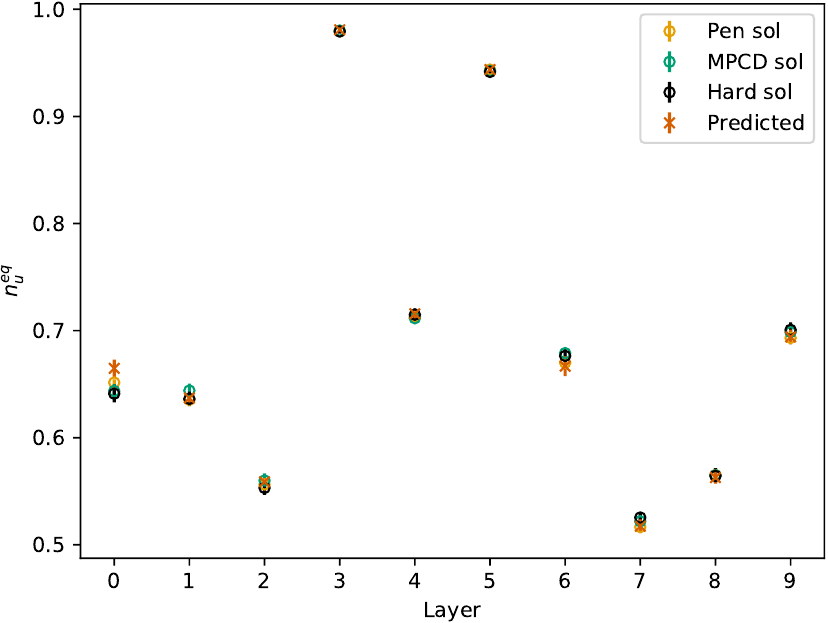}
  \caption{The $n_u^{\text{eq}}$ for crambin for all three solvent models at
  $\epsilon = 3.0$.}
  \label{fig:oldparamnu}
\end{figure}

Throughout each simulation, the fraction of the ensemble in the unbound state,
an estimator of the probability $n_u(t)$, was recorded at regular time
intervals. These data were then fit to an exponential of the form in
Eq.~(\ref{eq:solofode}), using a non-linear least squares procedure. From this
fit, the values of $k_{\text{r}}$ and $n_u^{\text{eq}}$ were obtained. Both the
analytical predictions and simulation results for the hard-sphere, MPCD, and
penetrating solvent models are plotted in Figs.~\ref{fig:oldparamkrates}
and~\ref{fig:oldparamnu}. The error bars in these figures, and in
Figs.~\ref{fig:ratios4},~\ref{fig:ratios5}, and~\ref{fig:newparamkrates}, were
obtained using bootstrap sampling of ensemble member
trajectories\cite{Efron:1982}. The number of bootstraps used for each pair of
error bars was $300$, and the size of each bootstrap sample was equivalent to
the size of the overall ensemble, $N_e$, typically larger than $10^{5}$. For
every set of bootstrap samples, the observed quantities, namely the rates
$k_{\text{r}}$, the equilibrium populations $n_u^{\text{eq}}$, as well as the
bond distance coefficient $D_{(ij)}$, were evaluated and used to establish $95
\%$ confidence intervals. To do so, the observed values were ordered from
smallest to largest, and the upper and lower confidence limits were the values
at the $2.5 \%$ and $97.5 \%$ percentiles, respectively. This procedure was
carried out since the data from the bootstraps were typically not normally
distributed.

For the diffusive Markov model of crambin, the predicted values of the
equilibrium probability $n_u^{\text{eq}}$ and the decay rate $k_r(\epsilon)$ for
a bond energy $\epsilon$ are given by
\begin{align}
    n_u^{\text{eq}} &= \frac{e^{-\epsilon} e^{\Delta S}}{1 + e^{-\epsilon}
    e^{\Delta S}}, \label{eq:crambinUnbondedProb} \\
    k_r(\epsilon) &= \frac{D_{(ij)} \left(1 + e^{-\epsilon} e^{\Delta S}
    \right)}{\overline{\tau}_{(ij)}^+ + e^{-\epsilon} e^{\Delta S} \,
    \overline{\tau}_{(ij)}^-} \label{eq:crambinRate}.
\end{align}
The confidence intervals for these predicted values depend on the statistical
errors of $D_{(ij)}$, $\Delta S$, and the mean first passage times. The error in
$\Delta S$ was obtained by running the dynamics of the same transition ten times
and finding the relative error across these results. The relative error of the
mean first passage times was obtained in the same way and was found to be $5 \%$
of the value of each mean first passage time or less in
Ref.~\onlinecite{Colberg:2022}, so for simplicity, here we set the relative
error to $5 \%$. This led to the total error for the predicted $k_{\text{r}}$
and state probability $n_u^{\text{eq}}$ being that indicated by the error bars
in the figures provided in the following section.

Since the solvent is in thermal equilibrium with the protein chain and the
instantaneous interaction solvent-monomer energies are always zero for all
solvent models considered here, the presence of the solvent does not influence
the equilibrium population of the unbound state $n_u^{\text{eq}}$. Hence, the
dynamics of the population of the unbound state for all solvent models must tend
to the same value, $n_u^{\text{eq}}$, a result confirmed in
Fig.~\ref{fig:oldparamnu}.

\section{RESULTS} \label{sec:results}

\begin{figure*}[htbp]
    \subfloat[
        \label{fig:layer3config}
        The bonding pattern for transition $1000000110$ to $1010000110$ in layer 3
    ]{%
        \includegraphics[height=7.0cm]{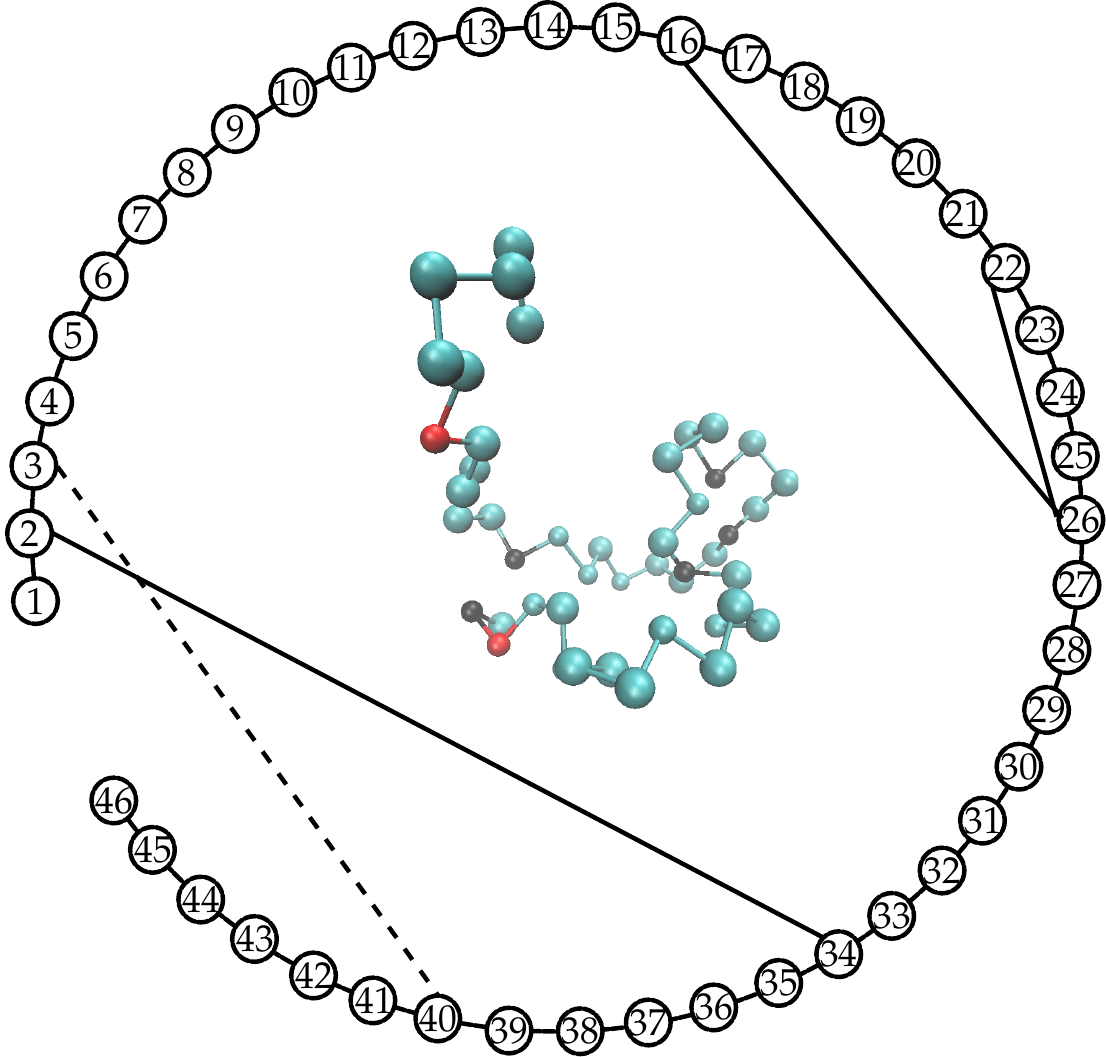}%
    }\hspace{20ex}
    \subfloat[
        \label{fig:layer8config}
        The bonding pattern for transition $1111001111$ to $1111101111$ in layer 8
    ]{%
        \includegraphics[height=7.0cm]{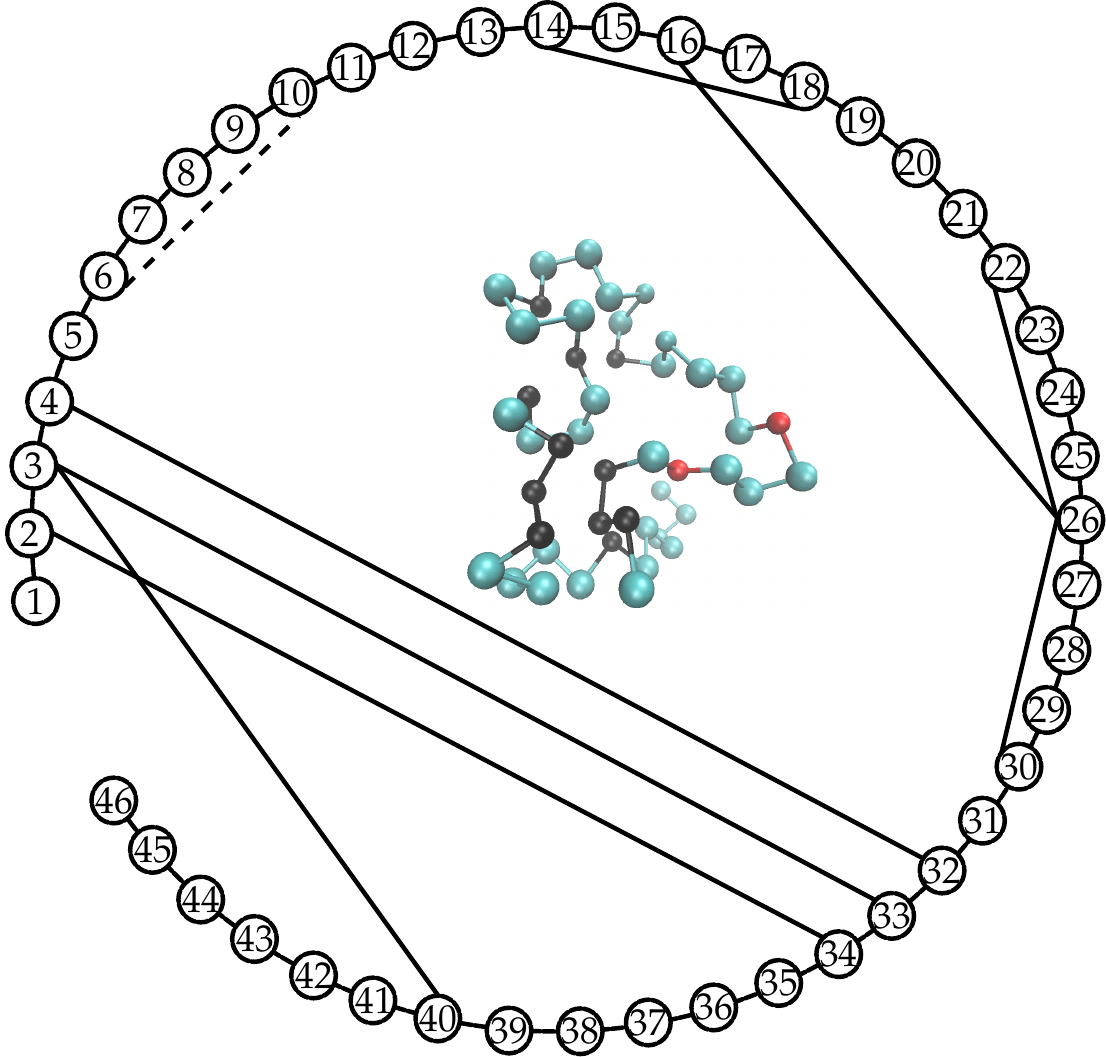}%
    }
    \caption{The bonding pattern for crambin in the transitions in layers 3 and
        8 represented as a linear chain, with the permanently maintained bonds
        shown as a solid link and the active bond as a dotted link. In the ball
        and stick models in the centers, the black beads are involved in
        permanent bonds, while the red beads are the transient bond beads.}
    \label{fig:layers3and8config}
\end{figure*}

Of the ten transitions occurring between the intermediate states of crambin
listed in Table~\ref{table:layers}, two representative layers with qualitatively
different characteristics containing three and eight permanent bonds (see
Fig.~\ref{fig:layers3and8config}), respectively, showcase the analysis of the
simulation data.

\begin{figure*}[htbp]
    \subfloat[
        \label{fig:configprob3}
        Configurational probability of the unbound state vs. time for transition
        $1000000110$ to $1010000110$ in layer 3
    ]{%
        \includegraphics[height=5.0cm]{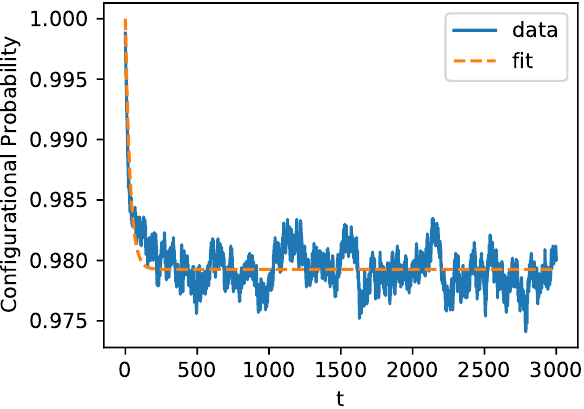}%
    }\hspace{25ex}
    \subfloat[
        \label{fig:msd3}
        MSD for transition $1000000110$ to $1010000110$ in layer 3
    ]{%
        \includegraphics[height=5.0cm]{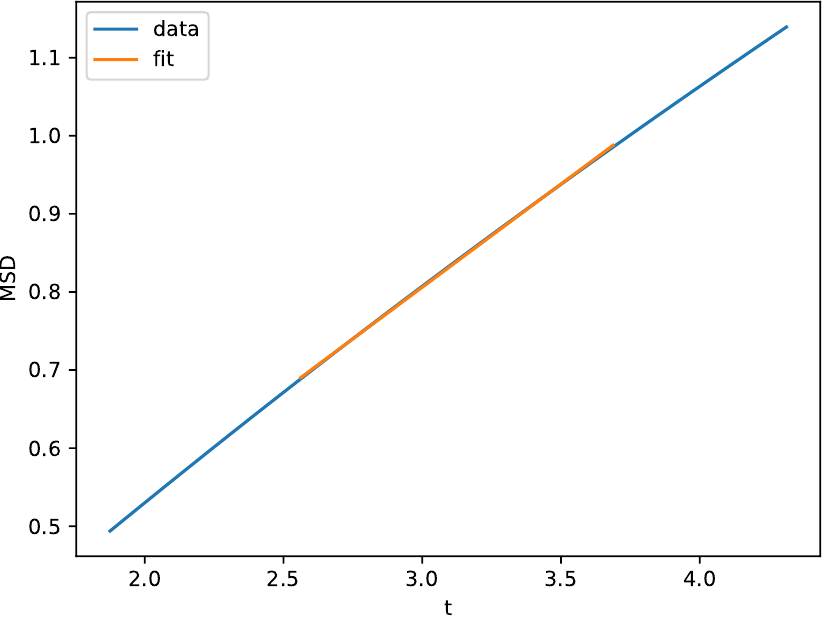}%
    } \\%
    \subfloat[
        \label{fig:configprob8}
        Configurational probability of the unbound state vs. time for transition
        $1111001111$ to $1111101111$ in layer 8
    ]{%
        \includegraphics[height=5.0cm]{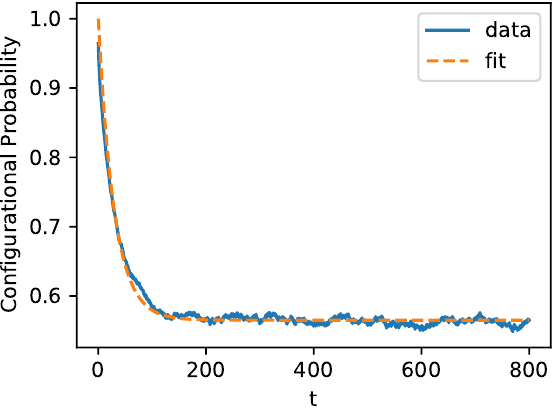}%
    }\hspace{25ex}
    \subfloat[
        \label{fig:msd8}
        MSD for transition $1111001111$ to $1111101111$ in layer 8
    ]{%
        \includegraphics[height=5.0cm]{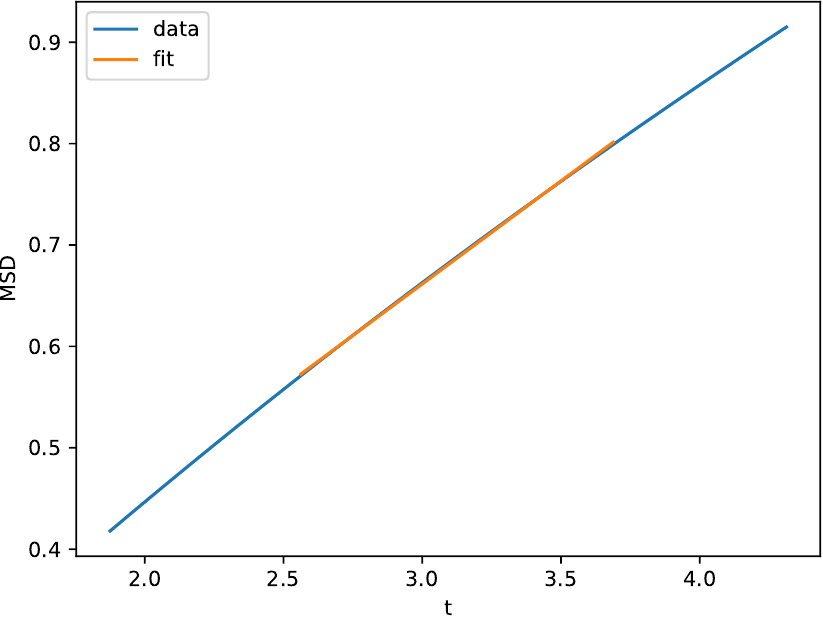}%
    }
    \caption{A comparison of two different layers in the presence of the
    penetrating solvent at $\epsilon = 3.0$ with an ensemble of $10,000$. The
    units for $t$ in this figure, and in Figs.~\ref{fig:layer3diffeps}
    and~\ref{fig:layer3diffsol} are in simulation time units $\tau$, where $\tau
    = \sqrt{m \ell^2/(k_B T)}$.}
    \label{fig:2layerspeneps3}
\end{figure*}

The detailed results for these
layers are presented in Fig.~\ref{fig:2layerspeneps3}. In layer 8, the active
bond is a fast-forming, short-range $\alpha$-helix bond occurring between beads
$[6, 10]$, whose configurational entropy difference was the second smallest of
the ten layers. In addition, the outer squared diffusion distance is short
($\overline{\tau}_{(ij)}^+ = 2.3 \, \ell^2$). In contrast, the active bond in
layer 3 is a slow-forming, long-range disulfide bridge between beads $[3, 40]$,
with a large mean first passage time (since $\overline{\tau}_{(ij)}^+ = 82 \,
\ell^2$) approximately 40 times larger than that of layer 8. The
configurational entropy difference $\Delta S = 6.97$ of this bond, which is
strongly correlated with the outer mean first passage time, is more than twice
as large as that of layer 8.

\begin{table}
    \begin{tabular}{c c c}
        \toprule
        Layer & $n_u^{\text{eq}}$ & $k_{\text{r}}$ \\
        \midrule
        3 (sol) & 0.9796 $\pm$ 5e-4 & 0.027 $\pm$ 6e-3 \\
        3 (MSM) & 0.9808 $\pm$ 7e-4 & 0.028 $\pm$ 2e-3 \\
        8 (sol) & 0.565 $\pm$ 3e-3 & 0.034 $\pm$ 2e-3 \\
        8 (MSM) & 0.563 $\pm$ 5e-3 & 0.034 $\pm$ 1e-3 \\
        \bottomrule
    \end{tabular}
    \caption{$n_u^{\text{eq}}$ and $k_{\text{r}}$ for layers $3$ and $8$ for the
    penetrating solvent and diffusive Markov state models at $\epsilon = 3.0$.}
    \label{table:Pbkratelayers38}
\end{table}

Looking at Fig.~\ref{fig:2layerspeneps3} and Table~\ref{table:Pbkratelayers38},
it is clear that while the equilibration rates $k_{\text{r}}$ for the two layers
are similar, the equilibrium populations $n_u^{\text{eq}}$ as well as the outer
first passage times are quite different. The value of $n_u^{\text{eq}}$ of layer
3 is much larger than that of layer 8, since the entropic cost of forming the
disulfide bond between the distant monomer pair $[3, 40]$ is high, even though a
$\beta$-sheet link between the monomer pair $[2, 34]$ already exists. In
comparison, the entropic cost of forming the local helical bond $[6, 10]$ in
layer 8, which decreases the potential energy of the molecule by the same factor
of $\epsilon = 3.0$, is lower ($\Delta S = 3.25$) since beads $6$ and $10$ are
restricted to being near one another by the eight other bonds already formed.
The same trend is seen across all other layers (see Fig.~\ref{fig:oldparamnu}).
In layer 5, the value of $n_u^{\text{eq}}$ is large because the bridge bond
being formed between beads $[16, 26]$ is relatively long-range (accordingly, the
configurational entropy difference is large: a value approximately twice that of
layer 8). Interestingly, we do not see this trend occurring in layers 4 and 7,
despite the active bonds in these layers being the $\beta$-sheet bond between
$[2, 34]$ in the former and the bridge bond between $[4, 32]$ in the latter. The
configurational entropy difference is small as well: the value for layer 8 is
approximately equivalent to layer 7 and three-fourths the magnitude of layer 4.
This is because in the unbound state, at least one long-range bond has been
formed, bringing the opposite ends of the protein together, making the formation
of subsequent long-range bonds in this region less entropically costly. In the
case of layers 4 and 7, the $\beta$-sheet bond between the beads $[3, 33]$ has
already been formed, facilitating the formation of future disulfide bridges and
$\beta$-sheet bonds in the vicinity of these two beads.

\begin{figure*}[htbp]
    \subfloat[
        \label{fig:configprob3eps4}
        Configurational probability of the unbound state vs. time at $\epsilon =
        4.0$
    ]{%
        \includegraphics[height=5.0cm]{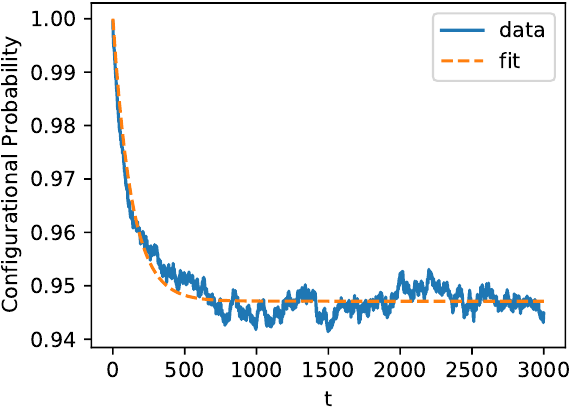}%
    }\hspace{25ex}
    \subfloat[
        \label{fig:configprob3eps5}
        Configurational probability of the unbound state vs. time at $\epsilon =
        5.0$
    ]{%
        \includegraphics[height=5.0cm]{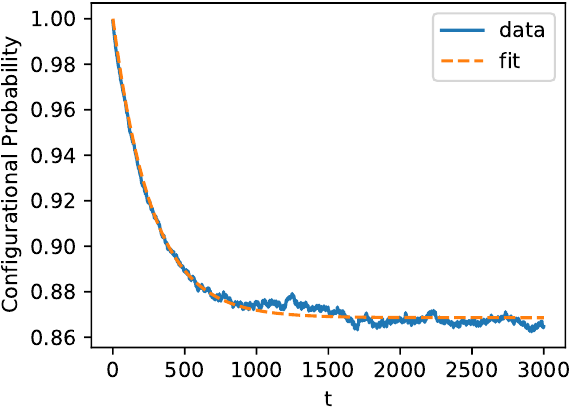}%
    }
    \caption{A comparison of two values of $\epsilon$ for layer $3$ in the
    presence of the penetrating solvent with an ensemble size of $N_e =
10,000$.}
    \label{fig:layer3diffeps}
\end{figure*}

\begin{figure*}[htbp]
    \subfloat[
        \label{fig:configprobmpcd}
        Configurational probability of the unbound state vs. time in the
        presence of the MPCD solvent with an ensemble size of $N_e = 10,000$
    ]{%
        \includegraphics[height=5.0cm]{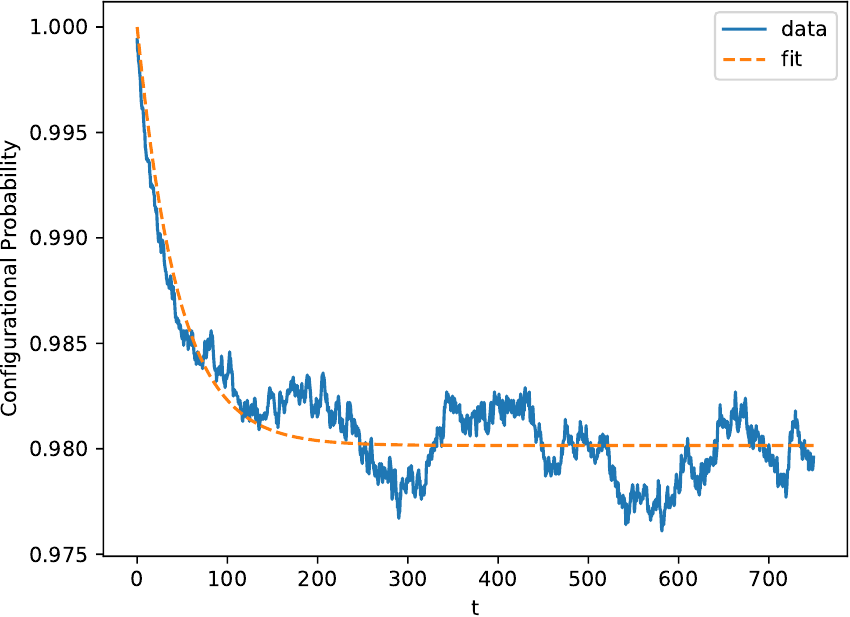}%
    }\hspace{25ex}
    \subfloat[
        \label{fig:configprobhard}
        Configurational probability of the unbound state vs. time in the
        presence of the hard-sphere solvent with an ensemble size of $N_e =
        4,000$
    ]{%
        \includegraphics[height=5.0cm]{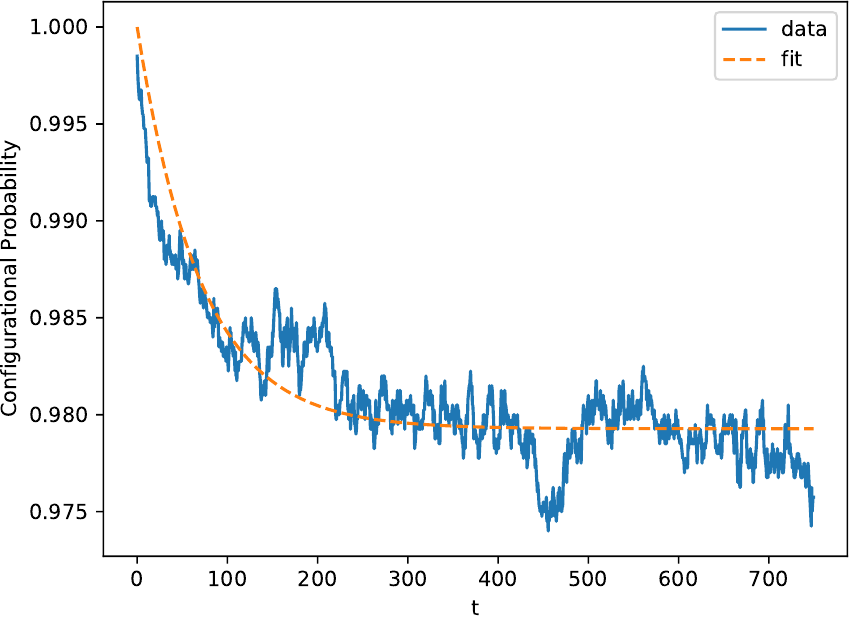}%
    }
    \caption{A comparison of the MPCD and hard-sphere solvent models for layer
    $3$ at $\epsilon = 3.0$.}
    \label{fig:layer3diffsol}
\end{figure*}

Another difference between layers $3$ and $8$ in Fig.~\ref{fig:2layerspeneps3}
is the magnitude of the fluctuations of the populations relative to the change
in the populations, $\delta n_u(0) = 1 - n_u^{\text{eq}} = n_b^{\text{eq}}$, of
the unbound state from the initial population of unity. When the
equilibrium-bound population $n_b^{\text{eq}}$ is small, larger ensembles are
required to resolve the relative population decay, $\delta n_u(t)/\delta
n_u(0)$, since the statistical uncertainty in the population with an ensemble
size of $N_e$ relative to $\delta n_u(0)$ is $\sigma_E/n_b^{\text{eq}} =
\left(n_u^{\text{eq}} / (N_e n_b^{\text{eq}})\right)^{1/2}$ when the populations
are Bernoulli distributed. Since the value of $n_b^{\text{eq}}$ increases with
the depth of the bonding well $\epsilon$, the statistical errors decrease for
larger values of $\epsilon$, and smaller ensembles can be used, as is apparent
in Fig.~\ref{fig:layer3diffeps}. The consideration of the relative magnitude of
fluctuations is important for computationally intensive solvent models such as
the hard-sphere model. The poor statistical resolution of the values of
$k_{\text{r}}$ and $n_u^{\text{eq}}$ for layer $3$ for the hard-sphere solvent
in Figs.~\ref{fig:oldparamkrates} and~\ref{fig:oldparamnu} is a direct
consequence of the limitations of the small effective ensemble size $N_e
n_b^{\text{eq}}$ for this system. This interpretation is confirmed by noting
that an increase in the value of $\epsilon$ noticeably reduces the relative
fluctuations, a consideration that is important for more detailed and
computationally expensive solvent models where the ensemble size $N_e$ is
limited (see Fig.~\ref{fig:layer3diffeps}).

\begin{figure}[htbp]
  \centering
  \includegraphics[width=\linewidth]{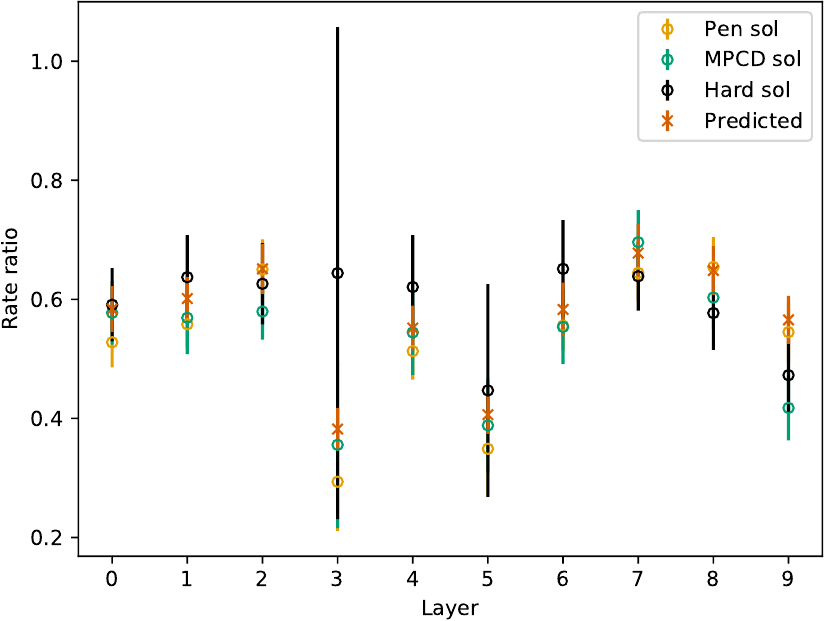}
  \caption{The rate ratios, $k_{\text{r}}(4)/k_{\text{r}}(3)$, for crambin for
  all three solvent models.}
  \label{fig:ratios4}
\end{figure}

\begin{figure}[htbp]
  \centering
  \includegraphics[width=\linewidth]{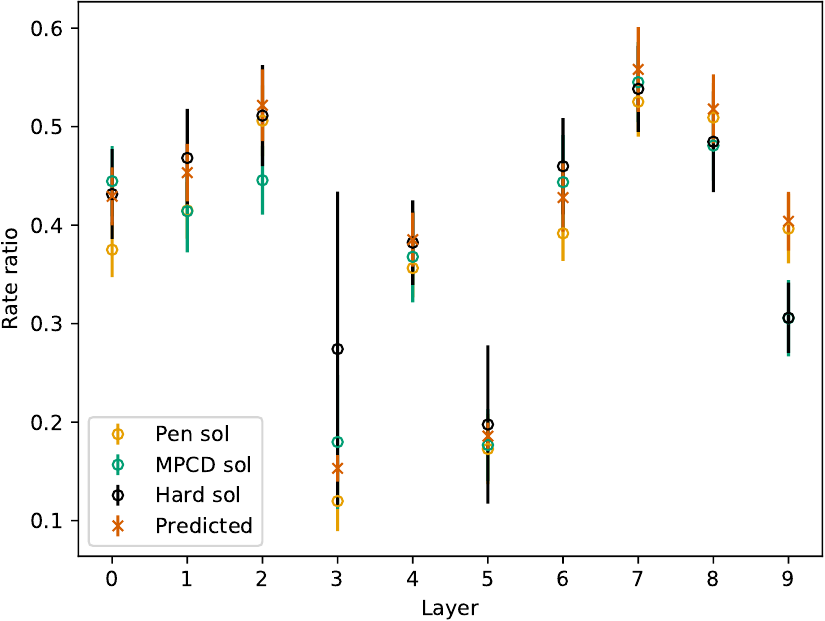}
  \caption{The rate ratios, $k_{\text{r}}(5)/k_{\text{r}}(3)$, for crambin for
  all three solvent models.}
  \label{fig:ratios5}
\end{figure}

The ratios between $k_{\text{r}}$ at different $\epsilon$ were calculated for
all three solvent models, and these are presented in Figs.~\ref{fig:ratios4}
and~\ref{fig:ratios5}. The rate ratios are computed as
\begin{align}
    \frac{k_{\text{r}}(\epsilon_1)}{k_{\text{r}}(\epsilon_0)} &= \left(\frac{1 +
    e^{-\epsilon_1} e^{\Delta S}}{1 + e^{-\epsilon_0} e^{\Delta S}}\right)
    \frac{\overline{\tau}_{(ij)}^+ + e^{-\epsilon_0} e^{\Delta S}
    \overline{\tau}_{(ij)}^-}{\overline{\tau}_{(ij)}^+ + e^{-\epsilon_1}
    e^{\Delta S} \overline{\tau}_{(ij)}^-} \nonumber \\
    &= \left(\frac{n_b^{\text{eq}}
    (\epsilon_0)}{n_b^{\text{eq}}(\epsilon_1)}\right)
    \frac{\overline{\tau}_{(ij)}^+ + e^{-\epsilon_0} e^{\Delta S}
    \overline{\tau}_{(ij)}^-}{\overline{\tau}_{(ij)}^+ + e^{-\epsilon_1}
    e^{\Delta S} \overline{\tau}_{(ij)}^-}. \label{eq:rateRatio}
\end{align}
Although the overall timescale of the relaxation rates $k_{\text{r}}$ is
determined primarily by the self-diffusion coefficients $D_{(ij)}$, the rates of
the different layers also depend on the outer squared diffusion distance
$\overline{\tau}_{(ij)}^+$. For large values of $\epsilon$ relative to the
entropic difference $\Delta S$, $k_{\text{r}}$ is inversely proportional to
$\overline{\tau}_{(ij)}^+$, a quantity that can be computed without explicitly
simulating the dynamics. By examining the ratio of rates at different values of
$\epsilon$, the statistical errors introduced by measuring the diffusion
coefficients can be bypassed, and the validity of the diffusive Markov state
model can be tested without reference to any dynamical quantities related to
time correlation functions.

Once again, the trend for the rate ratios predicted by Eq.~(\ref{eq:rateRatio})
is borne out in the simulations for all solvents. Since the ratio of the mean
first passage time contributions to the rate ratio is close to unity due to the
fact that $\overline{\tau}_{(ij)}^+ \gg \overline{\tau}_{(ij)}^-$, the ratio of
the bound populations primarily determines the change in rates as a function of
the well depth. The largest drop in the equilibration rate occurs for layers 3
and 5, where the ratio of the populations, $n_b^{\text{eq}}(3) /
n_b^{\text{eq}}(5) \ll 1$, decreases the most as $\epsilon$ increases. These
results confirm the importance of the equilibrium populations in the dynamics,
which are determined entirely by the geometric constraints, a non-trivial result
implied by the diffusive Markov state model.

\begin{figure}[htbp]
  \centering
  \includegraphics[width=\linewidth]{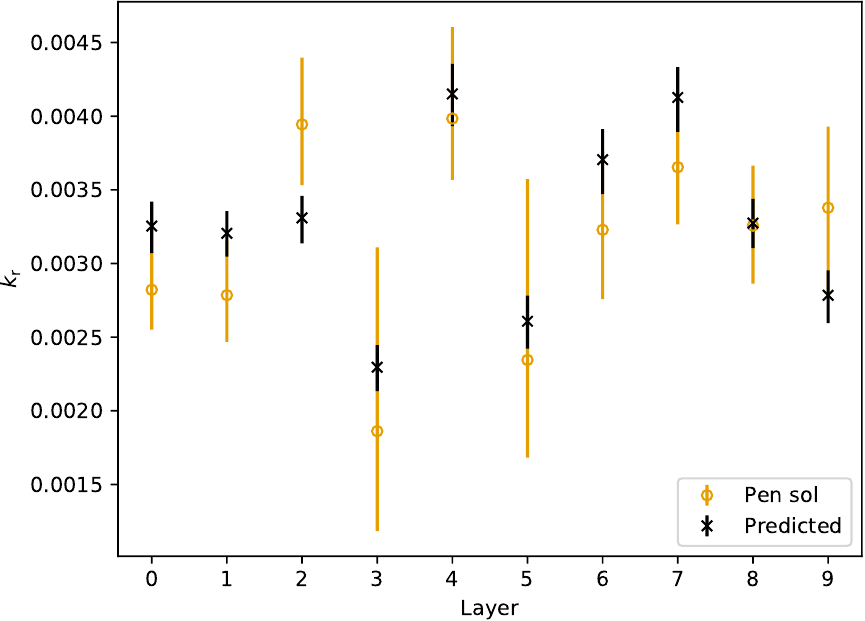}
  \caption{The relaxation rate $k_{\text{r}}$ for crambin for the penetrating
  solvent model at $\epsilon = 3.0$ with $N_e = 10^5$ when $\Delta t = 0.013
  \tau$ and the monomer diffusion coefficient is $D = 0.0045 \, \ell^2/\tau$.}
  \label{fig:newparamkrates}
\end{figure}

The simulation results described earlier apply to relatively dilute solvents
with larger bond distance diffusion coefficients than might apply to monomers of
proteins in crowded and dense environments. The relaxation rate $k_{\text{r}}$
for simulations of the penetrating solvent model with more frequent stochastic
rotations, where $\Delta t = 0.013 \tau$ and a monomer diffusion coefficient of
$0.0045 \, \ell^2/\tau$, is shown in Fig.~\ref{fig:newparamkrates}. The
equilibration dynamics lead to the same equilibrium probabilities of the unbound
state (data not shown). Unlike the equilibrium probability of the unbound state,
$n_u^{\text{eq}}$, which is independent of the dynamics, there is a clear
difference in the two sets of values of the decay rate $k_{\text{r}}$ when
comparing Fig.~\ref{fig:newparamkrates} to Fig.~\ref{fig:oldparamkrates}. As
predicted, the decay constant $k_{\text{r}}$ for each layer is considerably
smaller than under the previous simulation parameters since the diffusion
coefficient for a single bead is smaller by a factor of roughly $20$.
Nonetheless, the confidence intervals for the predicted rates, which differ from
the previous values by the ratio of their diffusion coefficients, and those
obtained from the simulations, overlap for all layers. The same trends in the
rates relative to the layer have been confirmed.

Simulations of more viscous MPCD and hard-sphere fluids are numerically
demanding, requiring prohibitively long computational times. For example, for
the more dilute solvent simulations of layer 2, where $D_{(ij)} = 0.03 \,
\ell^2/\tau$ and a fixed ensemble size of $10^5$ was used, the MPCD and
hard-sphere simulations required a factor of $44$ and $300$ times more
simulation time than the penetrating solvent simulations, respectively. The
relative computational demand of the different solvents is expected to be much
larger for more viscous fluids, particularly for the hard-sphere solvent, in
which the collision frequency of solvent particles scales quadratically with the
density.

\section{DISCUSSION AND CONCLUSIONS} \label{sec:concl}

In this work, we have confirmed that the diffusive Markov state model of a
linear chain of the crambin protein in solution accurately predicts the rates of
transitions between connected macrostates for three different solvent models and
different choices of bond energies (relative to the temperature). By
construction, the solvent models, which differ in the complexity and the
computational demand of their simulation by orders of magnitude, provide a
dissipative, fluctuating environment for the chain's monomers without
influencing the macrostates' probabilities. The simulation results suggest that
neither solvent-mediated hydrodynamic interactions between monomers nor the
structural ordering of the solvent particles are important factors in the
transition rates for the density of the solvents considered here. In the
low-density hard-sphere solvent (volume fraction of $0.04$), due to their small
sizes of the solvent radii ($0.1 \ell$) and of the monomer beads ($0.4 \ell$),
the radial distribution function at contact between the beads and the solvent,
$g(R) \sim 1$, is not large, and little structural effect is expected until much
higher volume fractions of hard-spheres. Solvent structural effects are not
anticipated to be important unless there is strong ordering in the solvent
around the protein chain and the solvent molecules are capable of stabilizing
configurations. The hydrodynamic effects are much more difficult to estimate
theoretically, even for a single solvated bead\cite{Hynes:1979}. The Stokes form
for the hydrodynamic friction, which can be derived for a large massive Brownian
particle whose $R \gg \xi$, where $\xi$ is the scale on which the radial
distribution varies, does not apply for the mesoscale-sized beads used in this
work to represent a coarse-grained amino acid\cite{Schofield:1992}. This
difficulty forces us to rely on numerical simulations of an isolated bead to
establish the equivalence of the self-diffusion coefficient of a monomer in each
of the solvents.

The only input from the equilibrium dynamics required in the diffusive Markov
state model is the bond distance diffusion coefficients, $D_{(ij)}$. This
transport property can be estimated from the rapid decay of the velocity
autocorrelation function or from linear fits of the short-time dynamics of the
mean-squared displacement. Both quantities do not require long trajectories and
are not computationally intensive, though their evaluation is complicated by
mode coupling and other solvent effects. Ideally, we would like to be able to
predict $D_{(ij)}$ from kinetic theory, but this is a daunting task since the
rapid local hard interactions between the nearest and next-nearest monomers in
the chain contribute to an internal friction that dominates over the friction
induced by monomer-solvent interactions since $D_{(ij)}$ differs substantially
from $2D_0$. Little kinetic theory work has been performed on models in which
particles are bound to one another through geometrical constraints, and how to
construct such a theory is an open question.

Another strong indication of the validity of the diffusion model is the
confirmation that the predicted ratios of the equilibration rates for different
bond energies, or, equivalently, the temperature, scale inversely with the ratio
of the equilibrium populations of the bound state $n_b^{\text{eq}}$ for the
respective bond energy $\epsilon$. The predicted rate ratios, which are
independent of the bond diffusion coefficient $D_{(ij)}$, are quantitatively
accurate for all layers and all three solvent models. Furthermore, the rates
obtained from the simulations themselves scale with $D_{(ij)}$, as predicted by
the diffusive Markov state model.

The success of the diffusive Markov state model for low-density solvents
suggests that the local interactions among monomers provide sufficient friction
for the motion of individual monomers, and the distances between them to be
diffusive in nature. The solvent and the chain are weakly coupled, and the
solvent-monomer interactions redistribute energy and effectively thermostat the
chain. Although it is debatable whether a lower or higher-density solvent is
more appropriate given the typical monomer size in the coarse-grained model, we
expect that increasing the solvent friction by increasing the density should not
alter the diffusive dynamics of monomers. This was confirmed for the penetrating
solvent model, for which the simulation of a solvated model protein is not
computationally prohibitive. However, higher-density solvents may allow for
dynamical, solvent-mediated correlations to perturb the diffusive dynamics of
the bonding distance.

In real solvents, such as water, solvent molecules can stabilize structures in a
protein via local hydrogen-bonding and electrostatic
interactions\cite{Bellissent-Funel:2016}. Although such interactions are not
included in the model, they could be incorporated by modeling solvent-monomer
interactions with square-well potentials, as is performed for the nonlocal
interactions among monomers in the chain. In such a model, the role of solvent
interactions in conformational changes\cite{Prakash:2012} could be analyzed
using the diffusive Markov state model. In the model studied here, hydrophobic
collapse\cite{Brylinski:2006,Lapidus:2007,Wirtz:2018} and other phenomena driven
by solvent-solvent interactions\cite{Sun:2022} can only be incorporated in a
mean-field fashion through configurationally-dependent site
energies\cite{Colberg:2022}.

As discussed previously\cite{Colberg:2022}, the simplicity of the temperature
and site energy dependence of the transition matrix in the diffusive Markov
state model enables not only the analysis of the mechanistic pathways between
two specified configurations and how this pathway changes with temperature but
also how site energies may be selected to optimize dynamical properties, such as
the average passage time between two chosen states. Since the rate ratios for
different choices of site energies are independent of $D_{(ij)}$, as has been
confirmed here by the molecular dynamics simulations, the optimization of site
energies can be carried out for a fixed set of values of the bond distance
diffusion coefficient.

\section*{ACKNOWLEDGMENTS}

Financial support from the National Sciences and Engineering Research Council of
Canada is gratefully acknowledged. Computations were performed on the Cedar
supercomputer at Simon Fraser University, which is funded by the Canada
Foundation for Innovation under the auspices of the Digital Research Alliance of
Canada, WestGrid, and Simon Fraser University. Code for this project is
available at \url{https://github.com/margaritacolberg}. The authors acknowledge
Vigneshwar Rajesh for assistance in coding and data analysis of the Markov state
model.

\appendix
\section{DECAY RATES AND FIRST PASSAGE TIMES}

For a two-state model in which a single bond between a pair of nonlocal beads
can form and break at a distance $r_c$, under certain conditions, the dynamics
of the non-equilibrium evolution of the population $n_u(t)$ of a configuration
$u$, given by the average of the indicator function $\mathbbm{1}_u(\bm{r})$ over
the non-equilibrium density $P(\bm{r},\bm{p},t)$ at the phase space point
$(\bm{r},\bm{p})$ of the linear chain of beads, can be reasonably well
approximated by a single exponential decay from an initial value of $n_u(0)$ to
an equilibrium value $n^{\text{eq}}_u$ of the form
\begin{equation} \label{eq:exponentialForm}
    \delta n_u(t) = n_u(t) - n^{\text{eq}}_u = \delta n_u(0) e^{-k_{\text{r}} t},
\end{equation}
where $k_{\text{r}}$ is the characteristic rate of decay. Here, we choose the
state $u$ to differ from a state $b$ by a single bond of length $r$, which is
formed by the passage between unbonded configurations $u$ that satisfy $r > r_c$
to bonded configurations satisfying $r < r_c$ while all other distance
constraints are met, so that
\begin{eqnarray} \label{eq:bondingCondition}
    n_{u}(t) &= \int^\prime d\bm{r} d\bm{p} \, H(r-r_c) P(\bm{r},\bm{p},t)
    \nonumber \\
    &= \int dr dp \, H(r-r_c) \rho(r,p,t).
\end{eqnarray}
In Eq.~\eqref{eq:bondingCondition}, $H(r-r_c)$ is the Heaviside function, and
the prime on the integral indicates that all other distance constraints are
maintained. Additionally, $\rho(r,p,t)$ is the probability density obtained by
integrating the full phase space density $P(\bm{r},\bm{p},t)$ over all but the
bonding degrees of freedom. Here, we assume that the linear chain of beads is
immersed in a stochastic environment due to the presence of either an effective
solvent (in the case of the penetrating model), a mesoscale solvent, or an
explicit solvent. If the solvent degrees of freedom evolve rapidly on the
timescale of the motion of the beads (for example, in the Brownian limit, in
which the mass ratio of the solvent and bead particles is $m/M \ll 1$), the
dynamics of the density of the reactive coordinate $r$ and its conjugate
momentum $p$ can be written as\cite{Mazur:1970}
\begin{equation} \label{eq:fp}
    \partial_t \rho(r,p,t) = {\cal L}_{\text{fp}} \, \rho(r,p,t),
\end{equation}
where ${\cal L}_{\text{fp}}$ is an effective Fokker-Planck operator
\begin{eqnarray} \label{eq:fpOperator}
    {\cal L_{\text{fp}}} &= -p/M \, \partial_r + \partial_r u(r) \, \partial_p +
    \gamma \partial_p \left(\partial_p + \beta p/M\right) \nonumber \\
    &+ \hat{T}_R \, \delta(r-r_c).
\end{eqnarray}
In Eq.~\eqref{eq:fpOperator}, $\hat{T}_R$ is a collision operator that accounts
for the step potential at $r=r_c$, in which the beads undergo an elastic
collision when $p_c > p > 0$ or experience an impulse that results in a
discontinuous jump in the momentum $p^\prime = p \pm p_c$, where $p_c =
\sqrt{\epsilon/(2M)}$ for a bond energy of $\epsilon$. In \eqref{eq:fpOperator},
$\gamma$ is the friction acting on the bonding beads that arises from the
interactions with the solvent and with other local beads, and $u(r)$ is the
continuous and smooth potential of mean force. In order for the population
dynamics to obey simple exponential behavior as in
Eq.~(\ref{eq:exponentialForm}), the solvent interactions must provide sufficient
dissipation to establish a separation of timescale between transitions between
states and the timescale of equilibration within each configuration.

The general solution of Eq.~(\ref{eq:fp}) is difficult but can be accomplished
by evaluating numerically the spectral decomposition of the Fokker-Planck
operator: The discontinuity at $r=r_c$ requires separating the solution into
spatial regions with $r<r_c$ and $r>r_c$ and matching the solutions at the
point of discontinuity. If one breaks the solution into two ``bulk" regions,
$\rho^{\pm}(r,p,t)$, which are spatially removed from the transition region, and
an interfacial region of width $2\sigma$,
\begin{align}
    \rho(r,p,t) &= \rho^s(r,p,t) + \rho^{+}(r,p,t) H(r-r_c-\sigma) \nonumber \\
    &\qquad + \rho^{-}(r,p,t) H(r_c-\sigma-r),
\end{align}
the dynamics of the density in the bulk regions is simple, while the dynamics of
$\rho^s(r,p,t)$ in the interfacial region is complicated by the effect of the
impulses that lead to a non-Boltzmann distribution of the
momentum\cite{Menon:1985,Kalinay:2012}. In the thin interfacial region
approximation corresponding to the limit $\sigma \rightarrow 0$, the solutions
of $\rho^{\pm}$ are extended to the transition state $r=r_c$. Using techniques
similar to the multipole expansions of excess surface densities in
hydrodynamics\cite{Ronis:1978,SchofieldKapral:2024}, in the thin layer
approximation, the effect of the metastable region can be replaced by a boundary
condition on the equations for the bulk densities,
$\rho^{\pm}(r,p,t)$\cite{SchofieldKapral:2024}.

In the high friction limit, where the distribution of momentum relaxes quickly
to the Boltzmann distribution in all spatial regions, one expects that the
dynamics of $\rho$ can be reduced to the Smoluchowski equation, in which the
marginal density $\rho(r,t) = \int d \bm{p} \, \rho(r,p,t)$ obeys the diffusion
equation\cite{Oppenheim:1990,Schofield:2014/095101},
\begin{equation} \label{eq:reactionDiffusion}
    \partial_t \rho(r,t)= {\cal L} \rho(r,t),
\end{equation}
where ${\cal L}$ is defined as
\begin{equation} \label{eq:smOperator}
    {\cal L} = D \, \partial_r e^{-\beta u(r)} \partial_r e^{\beta u(r)},
\end{equation}
and $D=(\beta^2 \gamma)^{-1}$. Continuity of probability leads to the jump
conditions\cite{Schofield:2014/095101}
\begin{align}
    \partial_r \rho(r,t) \big|_{r_c^{-}} &= \partial_r \rho(r,t) \big|_{r_c^+},
    \\
    e^{\beta u(r_c^+)} \rho(r_c^+,t) &= e^{\beta u(r_c^-)} \rho(r_c^{-},t).
\end{align}

For the diffusive system, the characteristic decay time $k_{\text{r}}^{-1}$ in
Eq.~(\ref{eq:exponentialForm}) can be expressed as a one-dimensional integral.
For a system with reflecting boundaries at $r=a$ and $r=b$,
\begin{align}
    \delta n_u(t) &= \int_a^b dr H(r-r_c) \left(\rho(r,t) - \rho_e(r)\right)
    \nonumber \\
    &= \int_a^b dr H(r-r_c) e^{{\cal L} t} {\cal Q}\rho(r,0) \\
    &= \int_a^b dr dr_0 \; \rho(r_0,0) H(r-r_c) e^{{\cal L} t} {\cal Q}
    \delta(r-r_0) \nonumber \\
    &= \int_a^b dr dr_0 \; \rho(r_0,0) H(r-r_c) e^{{\cal QL} t} {\cal Q}
    \delta(r-r_0), \nonumber
\end{align}
where $\rho_e(r)$ is the equilibrium density satisfying ${\cal L} \rho_e(r) =
0$, and ${\cal Q}$ is a projection operator that removes the projection onto the
equilibrium density\cite{Nadler:1985}, ${\cal Q} G(r) = G(r) - \rho_e(r)
\int_a^b dx G(x)$. From the Laplace transform $\delta \tilde{n}_u(z)$ of
Eq.~(\ref{eq:bondingCondition}), one has $k_{\text{r}}^{-1} = \lim_{z
\rightarrow 0} \delta \tilde{n}_u(z)/\delta n_u(0)$, hence
\begin{align}
    k_{\text{r}}^{-1} \delta n_u(0) &= \int_a^b dr dr_0 \; \rho(r_0,0) H(r-r_c)
    (-{\cal QL})^{-1} {\cal Q} \delta (r-r_0) \nonumber\\
    &= \int_a^b dr dr_0 \; \rho(r_0,0) H(r-r_c) \mu_{-1}(r,r_0), \label{eq:kInv}
\end{align}
where $\mu_{-1}(r,r_0) = -[{\cal QL} {\cal Q}]^{-1} \delta(r-r_0)$ or ${\cal
QL} \mu_{-1}(r,r_0) = -{\cal Q} \delta(r-r_0)$. Direct integration of this
equation gives
\begin{align}
    \mu_{-1}&(r,r_0) = D^{-1} J(r,r_0) - D^{-1} \rho_e(r) \int_a^b dr' \;
    J(r',r_0) \label{eq:mu-minus},
\end{align}
where
\begin{align}
    J&(r,r_0) = H(r_c-r) \rho_b(r) \int_r^{r_c} dy \; \frac{H(y-r_0) - n_b
    C_b(y)}{\rho_b (r)} \nonumber \\
    &+ H(r-r_c) \rho_u (r) \int_{r_c}^r dy \; \frac{H(r_0-y) - n_u
    (1-C_u(y))}{\rho_u (r)}. \nonumber
\end{align}
Here, $n_u^{\text{eq}} \rho_u(r) = H(r-r_c) \rho_e(r)$ defines the conditional
equilibrium density $\rho_u(r)$ in the unbonded state, $C_u(r) = \int_{r_c}^r dy
\, \rho_u(y)$ is the cumulative distribution of $\rho_u(r)$, and $\rho_b(r)$ and
$C_b(r)$ are similarly defined. Inserting Eq.~(\ref{eq:mu-minus}) into
Eq.~(\ref{eq:kInv}) gives
\begin{align} \label{eq:kInv-2}
    k_{\text{r}}^{-1} \delta n_u(0) &= \int_a^b dr_0 \; \rho(r_0,0)
    \left(n_b^{\text{eq}} \gamma_2(r_0) - n_u^{\text{eq}} \gamma_1(r_0)\right),
\end{align}
where
\begin{align}
    \gamma_1(r_0) &= D^{-1} \int_a^{r_c} J(r,r_0) \\
    &= \frac{H(r_c-r_0)}{D} \int_{r_0}^{r_c} dy \; \frac{C_b(y)}{\rho_b(y)} -
    n_b^{\text{eq}} \tau_b, \nonumber \\
    \gamma_2(r_0) &= D^{-1} \int_{r_c}^b J(r,r_0) \\
    &= \frac{H(r_0-r_c)}{D} \int_{r_c}^{r_0} dy \; \frac{1-C_u(y)}{\rho_u(y)} -
    n_u^{\text{eq}} \tau_u \nonumber,
\end{align}
and $\tau_b$ and $\tau_u$ are the inner and outer average first passage times
\begin{align}
    \tau_b &= \frac{1}{D} \int_a^{r_c} dr \; \frac{C_b(r)^2}{\rho_b(r)},
    \nonumber \\
    \tau_u &= \frac{1}{D} \int_{r_c}^{b} dr \; \frac{(1-C_u(r))^2}{\rho_u(r)}.
\end{align}
Finally, for the special case in which the initial value of $r_0$ is distributed
according to the conditional equilibrium distribution,
\begin{align} \label{eq:initialDistribution}
    \rho(r_0,0) &= \left[\frac{n_b(0)}{n^{\text{eq}}_b} H(r_c-r_0) +
    \frac{n_u(0)}{n^{\text{eq}}_u} H(r_0-r_c)\right] \rho_e(r_0),
\end{align}
we obtain
\begin{align} \label{eq:kInvResult}
    k_{\text{r}}^{-1} &= \mu_{-1}/\delta n_u(0) \\
    &= n_u^{\text{eq}} \tau_b + n_b^{\text{eq}} \tau_u \nonumber,
\end{align}
which is equivalent to Eq.~(\ref{eq:invKji}) in the main text.

In some cases, the decay of populations to equilibrium is not well-described by
a single exponential for diffusive processes. It is often found that although
the overall relaxation time is well-described by Eq.~(\ref{eq:kInvResult}), the
initial decay is faster and the later decay is slower than that predicted by
the single exponential approximation. If desired, more accurate relaxation
profiles can be constructed by higher-order Pad\'e
approximations\cite{Szabo:1980,Nadler:1983,Nadler:1985},
\begin{align}
    \delta{n_u}(t) / \delta n_u(0) &= \frac{1}{t_1 - t_2} \left(t_1 e^{-t/t_1} -
    t_2 e^{-t/t_2}\right) \nonumber \\
    &\qquad - \frac{\mu_1 t_1 t_2}{t_2 - t_1} \left(e^{-t/t_1} -
    e^{-t/t_2}\right).
\end{align}
Here, $\mu_1 = -\delta \dot{n}_u(0)$ is the initial rate of decay of $\delta
n_u(t)$, given by the average initial flux at the transition state position
$r=r_c$, and the timescale parameters satisfy
\begin{align}
    \mu_{-1}/\delta n_u(0) &= t_1 + t_2 - \mu_1 t_1 t_2 = n_u^{\text{eq}} \tau_b
    + n_b^{\text{eq}} \tau_u, \nonumber \\
    \mu_{-2}/\delta n_u(0) &= \int_0^\infty dt \; t \, \delta n_u(t)/\delta
    n_u(0) \nonumber \\
    &= t_1^2 + t_1 t_2 + t_2^2 - \mu_1 t_1 t_2(t_1 + t_2),
\end{align}
where
\begin{align}
    \mu_{-2} &= \int_a^b dr \; H(r-r_c) \mu_{-2}(r) \nonumber \\
    & = \delta n_u(0) \left(n_u^{\text{eq}}\tau_{2b} + n_b^{\text{eq}}
    \tau_{2u} - n_u^{\text{eq}} n_b^{\text{eq}} (\tau_b - \tau_u)^2\right),
\end{align}
and
\begin{align}
    \tau_{2b} &= \frac{1}{D^2} \int_a^{r_c} dr \, \rho_b(r) \left(\int_r^{r_c}
    dx \frac{C_b(x)}{\rho_b(x)}\right)^2, \nonumber \\
    \tau_{2u} &= \frac{1}{D^2} \int_{r_c}^{b} dr \, \rho_u(r)
    \left(\int_{r_c}^{r} dx \frac{1 - C_u(x)}{\rho_u(x)}\right)^2,
\end{align}
a result that can be obtained in a similar manner to that used to derive
Eq~(\ref{eq:mu-minus}) by integrating the equation ${\cal QL} \mu_{-2}(r) =
-\mu_{-1}(r)$ and using reflecting boundary conditions at the limits $a$ and
$b$.

We remark that all these results depend on several approximations: first, the
force exerted by the solvent on the beads is rapid on the timescale of the bead
motion. Second, the bead dynamics is overdamped and diffusive in all regions,
maintaining an equilibrium distribution of the relative momentum of the bonding
beads. Finally, there is a separation of timescale between the time required to
reach a conditional equilibrium in each of the configurations and the time
required to change from one bonding configuration to another.

\bibliography{JCP_solvent}

\end{document}